\RequirePackage{tikz}% otherwise there is a conflict with \usepackage{program} (in sn-jnl.cls) and tikz
\documentclass{sn-jnl}% Basic Springer Nature Reference Style/Chemistry Reference Style

\usepackage{overpic}%
\usepackage[nameinlink]{cleveref}%
\usepackage{amssymb}%
\usepackage{amsmath}%
\usepackage{mathtools}%
\usepackage{siunitx}% for SI-units
\DeclareSIUnit{\amperehour}{Ah}%
\usepackage{subcaption}% for subfloat and subfigure
\usepackage{xcolor}%

\usepackage{tikz}%
\usetikzlibrary{arrows,arrows.meta,bbox,calc,trees,positioning,patterns,patterns.meta,decorations.pathreplacing,shapes}%
\newcommand{\dx}[1]{\ensuremath{\,\mathrm{d}{#1}}}%
\newcommand*{\subsys}[1]{\text{\MakeUppercase{\romannumeral #1}}}
\newcommand{\sinew}[1]{\,\si{#1}}%
\makeatletter
\DeclareRobustCommand\widecheck[1]{{\mathpalette\@widecheck{#1}}}
\def\@widecheck#1#2{%
	\setbox\z@\hbox{\m@th$#1#2$}%
	\setbox\tw@\hbox{\m@th$#1%
		\widehat{%
			\vrule\@width\z@\@height\ht\z@
			\vrule\@height\z@\@width\wd\z@}$}%
	\dp\tw@-\ht\z@
	\@tempdima\ht\z@ \advance\@tempdima2\ht\tw@ \divide\@tempdima\thr@@
	\setbox\tw@\hbox{%
		\raise\@tempdima\hbox{\scalebox{1}[-1]{\lower\@tempdima\box
				\tw@}}}%
	{\ooalign{\box\tw@ \cr \box\z@}}}
\makeatother

\jyear{2022}%

\theoremstyle{thmstyleone}%
\theoremstyle{thmstyletwo}%

\theoremstyle{thmstylethree}%
\newtheorem{assumption}{Assumption}%

\begin{document}

\title[Resilient MPC of Distributed Systems Under Attack Using Local ADI]{Resilient Model Predictive Control of Distributed Systems Under Attack Using Local Attack Identification}

\author*[1]{\fnm{Sarah} \sur{Braun}}\email{sarah.braun@siemens.com}

\author[1]{\fnm{Sebastian} \sur{Albrecht}}\email{sebastian.albrecht@siemens.com}

\author[2]{\fnm{Sergio} \sur{Lucia}}\email{sergio.lucia@tu-dortmund.de}

\affil*[1]{\orgname{Siemens AG}, \orgaddress{\street{Otto-Hahn-Ring 6}, \postcode{81739} \city{M\"unchen}, \country{Germany}}}

\affil[2]{\orgname{TU Dortmund University}, \orgaddress{\street{August-Schmidt-Straße}, \postcode{44227} \city{Dortmund}, \country{State}}}

\abstract{
	With the growing share of renewable energy sources, the uncertainty in power supply is increasing. 
	In addition to the inherent fluctuations in the renewables, this is due to the threat of deliberate malicious attacks, which may become more prevalent with a growing number of distributed generation units.
	Also in other safety-critical technology sectors, control systems are becoming more and more decentralized, causing the targets for attackers and thus the risk of attacks to increase.
	It is thus essential that distributed controllers are robust toward these uncertainties and able to react quickly to disturbances of any kind. 
	To this end, we present novel methods for model-based identification of attacks and combine them with distributed model predictive control to obtain a resilient framework for adaptively robust control. 
	The methodology is specially designed for distributed setups with limited local information due to privacy and security reasons.
	To demonstrate the efficiency of the method, we introduce a mathematical model for physically coupled microgrids under the uncertain influence of renewable generation and adversarial attacks, and perform numerical experiments, applying the proposed method for microgrid control.
}


\maketitle

\section{Introduction}
\label{sec:intro}
Due to the energy transition, power generation is facing a technological change toward increasingly distributed generation, primarily from renewable energy sources.
Also in other technology areas such as industrial production or the transport sector, advancing automation and digitization are creating an increasing need for distributed control methods that can be applied to safety-critical systems in real time.
When designing such methods, it is important to take into account that distributed systems with many components can increase flexibility, but at the same time provide many targets for malicious attacks. 
Therefore, distributed control methods should be designed robustly and securely, and complemented with appropriate tools to increase the system's resilience to any type of disruption, which is particularly challenging in the event of unpredictable, adversarial attacks.

\emph{Model predictive control (MPC)} is one of the most popular control methods for dynamic systems in various fields of application as it applies to multivariable systems and allows to include constraints and cost functions in a natural way.
Based on updated measurements, it repeatedly computes optimal inputs to the system at each sampling time.
\emph{Distributed MPC (DMPC)} methods, see \cite{Christofides2013Distributed} for an overview and \cite{Arauz2021Cyber} for security-related DMPC, are designed for large systems of coupled subsystems and locally apply MPC in each subsystem.
In contrast to fully decentralized approaches where the neighbors' dynamic evolution is unknown to every subsystem, DMPC schemes involve some exchange of information among neighbors.
In \cite{Lucia2015Contract}, e.g., subsystems provide each other with corridors in which future values of their coupling variables lie. 
Given such information about the uncertainty range, \emph{robust MPC} can be applied to explicitly take uncertain influences into account when computing optimal inputs.
Robust MPC schemes typically build upon tube-based ideas as in \cite{Mayne2005Robust} or multi-stage approaches \cite{Lucia2013Multi}. 
It has been demonstrated in several works \cite{Wang2019Distributed,Braun2020Identifying,Braun2020Hierarchicala} that robust (D)MPC cannot only be applied for robustness against uncertain parameters or neighboring couplings, but also against adversarial attacks. 

While robust MPC can reduce the impact of disruptions if the uncertainty ranges are known, appropriate security measures for unknown attacks require that their presence and points of attack are recognized in the first place.
In this context, Pasqualetti et al.\ \cite{Pasqualetti2013Attack} introduce \emph{attack detection and identification (ADI)} as the tasks of revealing the presence of an attack and localizing all attacked system components. 
For both linear and nonlinear dynamics, there are many methods to detect and identify attacks or, closely related, unintentional system faults.
For a broad overview of physics- and control-based approaches we refer to the survey in \cite{Giraldo2018Survey}. 
Some works like \cite{Pasqualetti2013Attack,Boem2018Plug,Gallo2020Distributed} design unknown-input observers and employ one observer per attack scenario for identification, resulting in a combinatorial complexity.
Moreover, works on fault identification \cite{Boem2018Plug} often assume that all possible faults are known, which is an invalid assumption for adversarial attacks. 
In \emph{distributed} ADI, each subsystem employs its own estimator to detect and identify local perturbations, be it based on observer systems as in \cite{Boem2018Plug,Gallo2020Distributed,Boem2011Distributed} or sparse optimization problems as in \cite{Pan2015Online}. 
To represent the influence of other subsystems, the local problems typically involve measurements of the neighboring couplings transmitted by the neighbors \cite{Boem2018Plug} or approximated by adaptive local estimators \cite{Boem2011Distributed}.

In recent years, several approaches that intertwine the handling of attacks with (robust) DMPC have been published.
In \cite{Wang2019Distributed}, e.g., a DMPC-based strategy is presented by which systems reach resilient consensus even if some agents are malicious and transmit disturbed state values to their neighbors.
An attack identification method using Bayesian inference is introduced in \cite{Ananduta2020Resilient} and combined with DMPC to solve robust chance-constrained problems. 
The approach involves testing a series of hypotheses about the attack set and requires full enumeration of all possible attack scenarios.
To avoid the resulting combinatorial complexity, we combined a DMPC scheme from \cite{Lucia2015Contract} with our optimization-based global ADI method from \cite{Braun2021Attack} and proposed an adaptively robust DMPC method in \cite{Braun2021Adaptively} for targeted robust control against previously identified attack.

The contribution of this work, which is an extension of \cite{Braun2022Resilient}, consists in two novel approaches for distributed attack identification, a DMPC scheme embedding these ADI methods for adaptively robust control, and a numerical case study to illustrate the proposed resilient control framework using an example of interconnected microgrids under attack.
The new methods for model-based distributed ADI are derived in \Cref{sec:dadi} (significantly more detailed compared to \cite{Braun2022Resilient} and including one completely new method).
They involve a targeted exchange of information between neighbors and solve sparse optimization problems to locally identify an attack.
The identified insights are used by the DMPC framework for adaptively robust control presented in \Cref{sec:resilient dmpc} (considerably exceeding the summarized version in \cite{Braun2022Resilient}) to initiate suitable preparatory measures against previously identified attacks. 
Unlike the related technique introduced in \cite{Braun2021Adaptively}, it involves one of the new distributed ADI techniques presented in this paper.
Finally, we introduce here a more detailed numerical case study (in comparison to \cite{Braun2022Resilient}) with a nonlinear dynamic model for tertiary control of interconnected microgrids under attack in \Cref{sec:microgrid model} and perform numerical experiments with several attack scenarios in \Cref{sec:numerics}, illustrating the great potential of our resilient control framework for attacked microgrids with uncertain renewable generation.

\section{Problem Formulation}
\label{sec:problem}
We consider nonlinear dynamic systems with states $x \in \mathbb{X} \subseteq \mathbb{R}^{n_x}$, inputs $u \in \mathbb{U} \subseteq \mathbb{R}^{n_u}$, outputs $y \in \mathbb{Y} \subseteq \mathbb{R}^{n_y}$, and uncertain parameters $w \in \mathbb{W} \subseteq \mathbb{R}^{n_w}$ that behave according to discrete-time dynamics of the form
\begin{equation}
	\label{eq:global dynamic system discrete time}
	\begin{aligned}
		x^{k+1} &= f\left(x^k, u^k + a^k, w^k\right),\\
		y^{k+1} &= c\left(x^{k+1}\right),
	\end{aligned}
\end{equation}
with nonlinear functions $f: \mathbb{X} \times \mathbb{R}^{n_u} \times \mathbb{W} \rightarrow \mathbb{X}$ and $c: \mathbb{X} \rightarrow \mathbb{Y}$ that are assumed to be sufficiently smooth.
The system is exposed to the threat of potential attacks, which are modeled by attack inputs $a \in \mathbb{A}(u) \subseteq \mathbb{R}^{n_u}$ unknown to the controller. 
We consider arbitrary attack vectors $a$ and make no assumptions about the set $\mathbb{A}(u)$ of possible attacks.
While the attack model is additive in the input, an attack $a$ affects the states and outputs of the system in a nonlinear, nonadditive way.

The system is partitioned into a set $\mathcal{D}$ of subsystems $I$ with local states $x_I \in \mathbb{X}_I \subseteq \mathbb{R}^{n_{x_I}}$, local control inputs $u_I \in \mathbb{U}_I \subseteq \mathbb{R}^{n_{u_I}}$, local attack inputs $a_I \in \mathbb{A}_I(u) \subseteq \mathbb{R}^{n_{a_I}}$, local outputs $y_I \in \mathbb{Y}_I \subseteq \mathbb{R}^{n_{y_I}}$, and uncertain parameters $w_I \in \mathbb{W}_I \subseteq \mathbb{R}^{n_{w_I}}$.
A distributed version of the dynamic system in \labelcref{eq:global dynamic system discrete time} with local dynamic functions $f_I$ and local output functions $c_I$ is formulated as
\begin{equation}
	\label{eq:distributed dynamic system discrete time}
	\begin{aligned}
		x_I^{k+1} &= f_I\left(x_I^k, u_I^k + a_I^k, \widehat{z}_{\mathcal{N}_I}^k, w_I^k\right),\\
		z_I^{k+1} &= h_I\left(x_I^{k+1}\right),\\
		y_I^{k+1} &= c_I\left(x_I^{k+1}\right),
	\end{aligned}
\end{equation}
where the physical interconnection of subsystems is modeled through \emph{coupling variables} $z_I \in \mathbb{Z}_I \subseteq \mathbb{R}^{n_{z_I}}$ that are related to the local states $x_I$ through local coupling functions $h_I: \mathbb{X}_I \rightarrow \mathbb{Z}_I$. 
Since the dynamic evolution of the neighboring coupling variables $z_{\mathcal{N}_I}(t)$ during some time interval $t \in [t^k, t^{k+1}]$ is not determined by subsystem $I$, distributed models typically approximate~$z_{\mathcal{N}_I}(t)$ using some information provided by the neighbors. 
Here, we apply a parameterization scheme proposed in \cite{Kozma2014Distributed} and represent $z_I(t)$ on $[t^k, t^{k+1}]$ as the linear combination 
\begin{align*}
	z_I(t) = \sum_{j=1}^{\widehat{n}}z_I^{k,j}\beta_j^k(t)
\end{align*}
of $\widehat{n}$ basis functions $\beta_1^k, \dots, \beta_{\widehat{n}}^k: [t^k, t^{k+1}) \rightarrow \mathbb{R}$.
The coupling coefficients~$z_I^{k,j}$ are exchanged among neighbors and $\widehat{z}_I^k$ denotes the coefficient matrix \mbox{$\widehat{z}_I^k \coloneqq (z_I^{k,1}, \dots, z_I^{k,\widehat{n}}) \in \widehat{\mathbb{Z}}_I \subseteq \mathbb{R}^{n_{z_I}\times \widehat{n}}$.}
For a simplified notation, we introduce the chained local coupling function $\zeta_I \coloneqq h_I \circ f_I$ and the chained local output function $\eta_I \coloneqq c_I \circ f_I$. 
Similarly, the dense output coupling function $\widehat{\zeta}_I: \mathbb{X}_I \times \mathbb{R}^{n_u} \times \widehat{\mathbb{Z}}_{\mathcal{N}_I} \times \mathbb{W}_I \rightarrow \widehat{\mathbb{Z}}_I$ maps to the space  $\widehat{\mathbb{Z}}_I$ of coupling coefficients.

Based on the local coupling functions $\zeta_I$, so-called \emph{nominal} coupling values~$\bar{z}_I^{k}$ can be determined for the undisturbed case of no attack:
\begin{equation}
	\label{eq:definition nominal couplings}
	\bar{z}_I^{k+1} \coloneqq \zeta_I\left(x_I^k, u_I^k, \widehat{\bar{z}}^k_{\mathcal{N}_I}, 0\right).
\end{equation}
This nominal value is attained if no local attack is applied to the system, i.e., $a_I^k = 0$, no model uncertainty is present, i.e., $w_I^k = 0$, and all neighboring subsystems also behave according to their nominal values, i.e., $\widehat{z}^k_{\mathcal{N}_I} = \widehat{\bar{z}}^k_{\mathcal{N}_I}$.
For all methods presented in this paper we assume:
\begin{assumption}
	\label{ass:exchange of nominal coupling values}
	At each time $k$, each subsystems $I\in \mathcal{D}$ transmits the predicted nominal values $\widehat{\bar{z}}_I^k, \dots, \widehat{\bar{z}}_I^{k+N_{\text{p}}-1}$ of its coupling coefficients with prediction horizon $N_{\text{p}} \in \mathbb{N}$ to its neighbors.
\end{assumption}
Given this exchange of information among neighbors, the above definition in \labelcref{eq:definition nominal couplings} allows for a distributed calculation of the nominal values in a receding horizon fashion, where the local values computed and transmitted by subsystem $I$ at time $k$ are used by its neighbors to update their predictions one time step later. 
The definition further requires suitable initial values $\widehat{\bar{z}}_I^0$ to be available. 
For simplicity, we assume the system to be in a steady state $x^0$ at time~$0$ and take $\bar{z}_I^{0,j}=h_I(x_I^0)$ for all $j \in \{1, \dots, \widehat{n}\}$. 

Finally, each subsystem is subject to a set of local constraints 
\begin{align}
\label{eq:local constraints}
	g_I\left(x_I^k, u_I^k + a_I^k, \widehat{z}_{\mathcal{N}_I}^k, w_I^k\right) \leq 0
\end{align}
for some nonlinear function $g_I: \mathbb{X}_I \times \mathbb{R}^{n_{u_I}}  \times \widehat{\mathbb{Z}}_{\mathcal{N}_I} \times \mathbb{W}_I \rightarrow \mathbb{R}^{n_{g_I}}$ that must be satisfied at all times.

\section{Distributed Attack Identification Based on Sparse Optimization}
\label{sec:dadi}
The goal of this section is to propose a distributed ADI method that, in contrast to global methods, does not involve a central authority which has access to a global model of the system.
Instead, we formulate a bank of local problems that allow each subsystem to identify a suspicion $a^{\ast}_I$ about a potential local attack $a_I$ based on locally available model knowledge and, possibly, interaction with its neighboring subsystems.
In contrast to the centralized ADI method we presented in \cite{Braun2021Attack}, no local model knowledge is published globally.

Before that, we briefly recall the distributed method for the \emph{detection} of attacks that has already been presented in \cite{Braun2021Attack}. It is based on each subsystem~$I$ monitoring the deviations $\Delta z_I^{k+1} \coloneqq z_I^{k+1} - \bar{z}_I^{k+1}$ in its local coupling variables from the respective nominal values $\bar{z}_I^{k+1}$.
As the nominal values $\bar{z}_I^{k+1}$ defined in \labelcref{eq:definition nominal couplings} are attained in the undisturbed case, a deviation from them indicates a disturbance at time $k$.
Using a detection threshold $\tau_{\text{D}} \in \mathbb{R}_{> 0}$, the method detects an attack if $\|\Delta z_I^{k+1}\|_{\infty} > \tau_{\text{D}}$ for any $I$, i.e., if a distinct deviation is observed in any subsystem.
To ensure that only significant attacks are revealed rather than small model inaccuracies or measurement noise, one can assume a probability distribution of the uncertainty and define $\tau_{\text{D}}$ accordingly as in, e.g., \cite{Boem2018Plug}. 
Even if subsystem $I$ detects an attack by observing a clear deviation $\|\Delta z_I^{k+1}\|_{\infty} > \tau_{\text{D}}$, it does not necessarily have to be caused by an attack $a_I^k \neq 0$ in $I$, but can just as well be caused by neighboring subsystems deviating from their nominal couplings $\widehat{\bar{z}}_{\mathcal{N}_I}^k$.
Identifying the root of the disturbance and thus locating the attack is the task of attack \emph{identification}.

\begin{figure}[tb]%
	\centering%
	\begin{overpic}[width=0.93\linewidth]{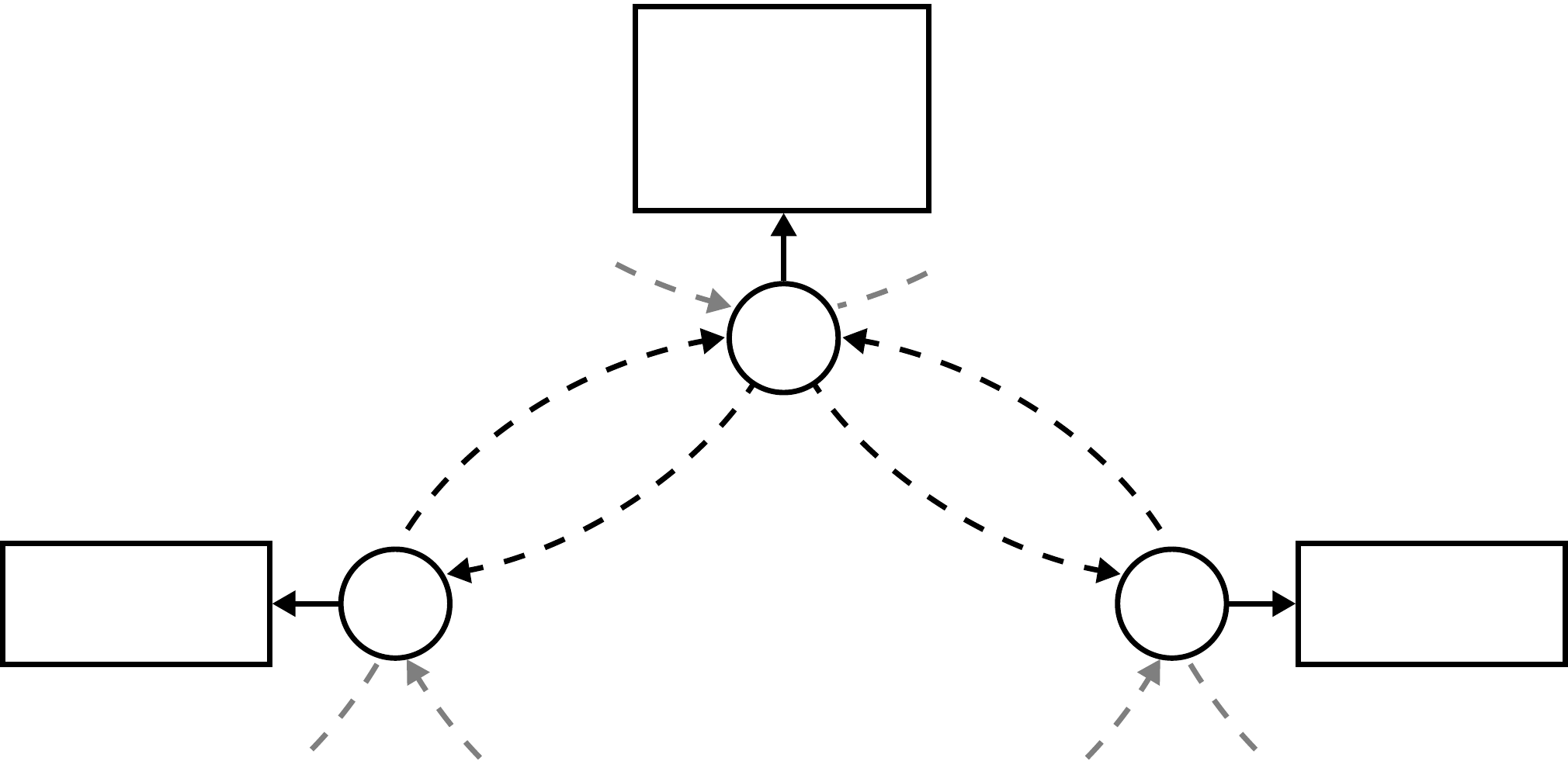}%
		% Subsystems
		\put(24.5,9.){I}%
		\put(73.9,9.){L}%		
		\put(48.7,26.){K}%
		% Local ADI
		\put(1.1,9){\small\textbf{Local ADI}}%
		\put(83.9,9){\small\textbf{Local ADI}}%		
		\put(36.8,40.3){%
			\begin{minipage}{3cm}
				\centering
				\small\textbf{Local ADI} \\
				\small involving \\
				\small problem \labelcref{opt:identification problem local}
			\end{minipage}%
		}%
		% Information Subsys I
		\put(27,21.3){\rotatebox{30}{\textcolor{blue}{$\widehat{\bar{z}}_I^k$, $\Delta \widehat{z}_I^k$}}}%
		% Information Subsys K
		\put(35,9.8){\rotatebox{30}{\textcolor{blue}{$\widehat{\bar{z}}_K^k$, $\Delta \widehat{z}_K^k$}}}%
		\put(53,15.6){\rotatebox{-30}{\textcolor{blue}{$\widehat{\bar{z}}_K^k$, $\Delta \widehat{z}_K^k$}}}%
		% Information Subsys L	
		\put(62,26.1){\rotatebox{-30}{\textcolor{blue}{$\widehat{\bar{z}}_L^k$, $\Delta \widehat{z}_L^k$}}}%
	\end{overpic}%
	\caption{If neighboring subsystems in a distributed system exchange suitable information about their local coupling variables, each subsystem can employ a local ADI method to identify suspicions about unknown local attack inputs.}
	\label{fig:exchange adi information distributed}
\end{figure}
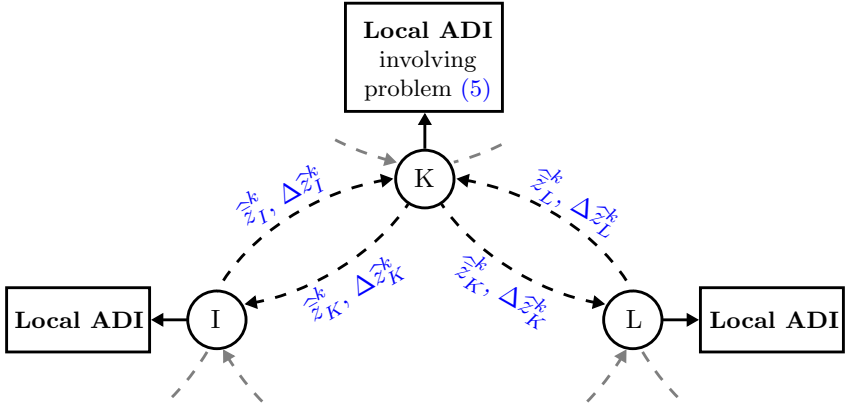

In this paper, also the identification of attacks is addressed in a distributed manner.
Depending on the amount and type of information that neighbors are willing to share, we derive two different versions of local identification problems. 
Clearly, the more specific the transmitted information describes the neighbors' behavior, the more precisely a local attack or even an attack on neighboring subsystems can be identified.
Therefore, the design of a local identification problem needs to suitably balance the required amount of information and the significance of the obtained suspicions.
For the first local identification problem that we establish, we propose that in addition to the exchange of nominal values $\widehat{\bar{z}}_I^k$ according to \Cref{ass:exchange of nominal coupling values}, also the deviations $\Delta \widehat{z}_I^k$ in the coupling coefficients are repeatedly transmitted to neighboring subsystems. 
This exchange is performed at each step $k$ when an attack is detected and is illustrated in \Cref{fig:exchange adi information distributed}.
Assuming that each subsystem can locally measure the impact onto its output variables $y_I^{k+1} \in \mathbb{Y}_I \subseteq \mathbb{R}^{n_{y_I}}$, we formulate a local attack identification problem to identify local attacks $a_I^k$ as
\begin{equation} 
	\label{opt:identification problem local}
	\begin{aligned}
		&\min_{a_I} &&\left\|a_I\right\|_1 ~~\text{ s.t.}	\\
		&&&\left\|y_I^{k+1} - \eta_I\left(x_I^k, u_I^k + a_I, \widehat{\bar{z}}_{\mathcal{N}_I}^k + \Delta \widehat{z}_{\mathcal{N}_I}^k, 0\right)\right\|_2
		\leq 
		\varepsilon_I.
	\end{aligned}
\end{equation}
A solution of problem \labelcref{opt:identification problem local}, which has already been proposed in \cite{Braun2022Resilient}, identifies a local suspicion $a^{\ast}_I$ for some subsystem $I$, which is $\ell_1$-norm sparsest among all possible attack vectors in $\mathbb{R}^{n_{u_I}}$ that explain the observed output $y_I^{k+1}$ according to the local model with output function $\eta_I$ up to a predefined tolerance $\varepsilon_I \in \mathbb{R}_{\geq 0}$, neglecting possible parametric uncertainties $w_I^k$.
While the optimization variable $a_I \in \mathbb{R}^{n_{u_I}}$ represents the unknown attack to be identified, the local state $x_I^k$, input $u_I^k$, and output $y_I^{k+1}$ are measured or known from local control computations, and the values $\widehat{\bar{z}}_{\mathcal{N}_I}^k$ and $\Delta \widehat{z}_{\mathcal{N}_I}^k$, and thus the actual neighboring coupling values $\widehat{z}_{\mathcal{N}_I}^k = \widehat{\bar{z}}_{\mathcal{N}_I}^k + \Delta \widehat{z}_{\mathcal{N}_I}^k$, are transmitted by neighbors.
Computing a \emph{sparse} suspicion to identify the attack is common in related work on attack identification, e.g., \cite{Pasqualetti2013Attack,Pan2015Online} and is justified by the observation that attackers typically have limited resources and are thus confined to impairing only few control components.
Some approaches formulate related optimization problems using an $\ell_0$-``norm'' cost term $\|a_I\|_0$ to count the number of attacked inputs, but solving them requires solution methods from mixed integer programming and is NP-hard \cite{Pasqualetti2013Attack}. 
To reduce the computational complexity and to obtain a numerically more tractable problem, the $\ell_0$-``norm'' is typically relaxed by the $\ell_1$-norm, see also \cite{Braun2021Attack,Candes2005Decoding}. 

If the neighboring subsystems in $\mathcal{N}_I$ agree to provide $I$ with even more information, subsystem $I$ can apply another version of local identification problem, which allows to draw not only conclusions about a potential local attack $a_I^k$, but even about attack inputs $a_{\mathcal{N}_I}^k$ in the neighborhood of $I$.
Since distributed methods are often applied when sensitive local information must not be made publicly available, we assume that neighbors still seek to keep their analytical model knowledge private and are only willing to reveal suitable numerical derivative information evaluated at the current iterate.
We pursued a similar approach for the centralized ADI method presented in \cite{Braun2021Attack}, involving the exchange of locally computed sensitivity matrices.
To motivate which kind of sensitivity information about the dynamic behavior of its neighbors subsystem~$I$ requires, we approximate the neighboring influence onto the local output $y_I$ by a first-order Taylor expansion of $\eta_I(x_I^k, u_I^k + a_I^k, \widehat{z}_{\mathcal{N}_I}^k, 0)$ in the $\widehat{z}_{\mathcal{N}_I}$-argument around the nominal value $\widehat{\bar{z}}_{\mathcal{N}_I}^k$.
To this end, we define a local sensitivity function $S^z_{I\mathcal{N}_I}: \mathbb{R}^{n_{u_I}} \rightarrow \mathbb{R}^{n_{y_I} \times n_{z_{\mathcal{N}_I}}}$, which maps each given attack input $a_I \in \mathbb{R}^{n_{u_I}}$ to the Jacobian
\begin{align*}
	S^z_{I\mathcal{N}_I}\left(a_I\right) 
	\coloneqq
	\frac{\partial \eta_I}{\partial \widehat{z}_{\mathcal{N}_I}}\left(x_I^k, u_I^k + a_I, \widehat{\bar{z}}_{\mathcal{N}_I}^k, 0\right),
\end{align*} 
that expresses the first-order dependence of the local output function~$\eta_I$ on the neighboring coupling variables $\widehat{z}_{\mathcal{N}_I}$.
It can be evaluated locally by $I$ and allows to approximate the local output variables $y_I^{k+1}$ according to Taylor's theorem, e.g., \cite[§7]{Forster2010Analysis} as
\begin{align}
	\label{eq:taylor expansion for distributed adi}
	y_I^{k+1} 
	=
	\eta_I\left(x_I^k, u_I^k + a_I^k, \widehat{\bar{z}}_{\mathcal{N}_I}^k, 0\right)
	+
	S^z_{I\mathcal{N}_I}\left(a_I^k\right) \Delta \widehat{z}_{\mathcal{N}_I}^k
	+
	R^{\text{lin}}_I
	+
	R^w_I.
\end{align}
Here, the remainder term of the Taylor expansion is denoted by $R^{\text{lin}}_I$ and can be estimated similar to the upper bound proven in \cite{Braun2021Attack}.
The term $R^w_I$ represents a model error which occurs as all uncertain parameters $w_I^k$ are considered zero in \labelcref{eq:taylor expansion for distributed adi} and due to the fact that the distributed model in \labelcref{eq:distributed dynamic system discrete time} only approximates the global dynamics in \labelcref{eq:global dynamic system discrete time}.

At this point, the additional sensitivity information provided by the neighbors $\mathcal{N}_I$ of $I$ comes into play. 
Denoting the coupling coefficients of the neighbors' neighbors by $\widehat{z}_{\mathcal{N}_{\mathcal{N}_I}}$, we introduce two types of sensitivity matrices as
\begin{align*}
	\widehat{S}^a_{\mathcal{N}_I}
	\coloneqq
	\frac{\partial \widehat{\zeta}_{\mathcal{N}_I}}{\partial a_{\mathcal{N}_I}}\left(x_{\mathcal{N}_I}^k, u_{\mathcal{N}_I}^k, \widehat{\bar{z}}_{\mathcal{N}_{\mathcal{N}_I}}^k, 0\right)
	~\text{ and }~
	\widehat{S}^z_{\mathcal{N}_I}
	\coloneqq
	\frac{\partial \widehat{\zeta}_{\mathcal{N}_I}}{\partial \widehat{z}_{\mathcal{N}_{\mathcal{N}_I}}}\left(x_{\mathcal{N}_I}^k, u_{\mathcal{N}_I}^k, \widehat{\bar{z}}_{\mathcal{N}_{\mathcal{N}_I}}^k, 0\right).
\end{align*}
The function $\widehat{\zeta}_{\mathcal{N}_I}$ denotes the dense coupling function of all neighbors in $\mathcal{N}_I$, which maps to the space $\widehat{\mathbb{Z}}_{\mathcal{N}_I}$ of coupling coefficients $\widehat{z}_{\mathcal{N}_I}$ and is obtained by combining the local dense coupling functions $\widehat{\zeta}_{L}$ for all $L \in \mathcal{N}_I$.
Hence, the sensitivity matrices $\widehat{S}^a_{\mathcal{N}_I}$ and $\widehat{S}^z_{\mathcal{N}_I}$ represent first-order approximations of how disturbances in $u_{\mathcal{N}_I}$ and $\widehat{z}_{\mathcal{N}_{\mathcal{N}_I}}$ affect the coupling coefficients $\widehat{z}_{\mathcal{N}_I}$.
If the neighbors in $\mathcal{N}_I$ provide subsystems $I$ with this information, the deviation $\Delta \widehat{z}_{\mathcal{N}_I}^k$ of neighboring couplings $\widehat{z}_{\mathcal{N}_I}^k$ from their transmitted nominal values $\widehat{\bar{z}}_{\mathcal{N}_I}^k$ can be expressed as 
\begin{align}
	\label{eq:expression for delta z neighbors}
	\Delta \widehat{z}_{\mathcal{N}_I}^k 
	= 
	\widehat{S}^a_{\mathcal{N}_I}
	a_{\mathcal{N}_I}^k
	+
	\widehat{S}^z_{\mathcal{N}_I}
	\Delta \widehat{z}_{\mathcal{N}_{\mathcal{N}_I}}^k
	+ 
	R^{\text{lin}}_{\mathcal{N}_I}
	+
	R^w_{\mathcal{N}_I}.
\end{align}
The model error $R^w_{\mathcal{N}_I}$ is caused by the uncertain influence of the parameters $w_{\mathcal{N}_I}^k$ and the linearization error $R^{\text{lin}}_{\mathcal{N}_I}$ denotes the Taylor remainder term when expanding the neighbors' coupling function $\widehat{\zeta}_{\mathcal{N}_I}$ around  $\widehat{\bar{z}}_{\mathcal{N}_{\mathcal{N}_I}}^k$.
%Both error terms can be estimated similar as discussed above for $R^w_I$ and $R^{\text{lin}}_I$.
The representation in \labelcref{eq:expression for delta z neighbors} gives subsystem $I$ more detailed insights into why its neighbors' coupling values $\widehat{z}_{\mathcal{N}_I}^k$ differ from the nominal values $\widehat{\bar{z}}_{\mathcal{N}_I}^k$. 
More precisely, it allows subsystem $I$ to distinguish whether the deviation is caused by an attack $a_{\mathcal{N}_I}^k$ that the neighbors are exposed to or whether they pass on the disturbing effect of any of their neighbors.
In order to figure out which source of disturbance applies, subsystem $I$ solves the following local identification problem with optimization variables $a_I$, $a_{\mathcal{N}_I}$, and $\Delta \widehat{z}_{\mathcal{N}_{\mathcal{N}_I}}$:
\begin{equation} 
	\label{opt:identification problem distributed}
	\begin{aligned}
		&\min_{a_I,a_{\mathcal{N}_I},\Delta \widehat{z}_{\mathcal{N}_{\mathcal{N}_I}}}
		~~\left\|a_I\right\|_1
		+ \left\|a_{\mathcal{N}_I}\right\|_1
		+ \left\|\Delta \widehat{z}_{\mathcal{N}_{\mathcal{N}_I}}\right\|_1 ~~\text{ s.t.}\\
		&\Big\|y_I^{k+1} - \eta_I\left(x_I^k, u_I^k + a_I, \widehat{\bar{z}}_{\mathcal{N}_I}^k, 0\right)
		+ S^z_{I\mathcal{N}_I}(a_I) \left(\widehat{S}^a_{\mathcal{N}_I} 	a_{\mathcal{N}_I} + \widehat{S}^z_{\mathcal{N}_I}\Delta \widehat{z}_{\mathcal{N}_{\mathcal{N}_I}}\right)\Big\|_2 \leq \varepsilon_I.
	\end{aligned}
\end{equation}
An optimal solution $(a^{\ast}_I, a^{\ast}_{\mathcal{N}_I}, \Delta \widehat{z}^{\ast}_{\mathcal{N}_{\mathcal{N}_I}})$ of problem \labelcref{opt:identification problem distributed} is sparsest with respect to the $\ell_1$-norm among all feasible points satisfying the constraints, which are obtained by combining \labelcref{eq:taylor expansion for distributed adi,eq:expression for delta z neighbors} and neglecting all error terms.
Similar to problem \labelcref{opt:identification problem local}, the constraints are relaxed by some tolerance $\varepsilon_I \in \mathbb{R}_{\geq 0}$ to account for model inaccuracies.
Besides the local quantities $u_I^k$, $y_I^{k+1}$, and $x_I^k$, which are known, measured, or estimated by the local control scheme, problem \labelcref{opt:identification problem distributed} also involves the nominal coefficients $\widehat{\bar{z}}_{\mathcal{N}_I}^k$, which are assumed to be exchanged among neighboring subsystems according to \Cref{ass:exchange of nominal coupling values}. 
Instead of the coupling deviations $\Delta \widehat{z}_{\mathcal{N}_I}^k$, the exchange of which is illustrated in \Cref{fig:exchange adi information distributed} and taken for granted by the first local identification problem \labelcref{opt:identification problem local}, the new distributed ADI approach requires all neighbors to provide the sensitivity matrices $\widehat{S}^a_{\mathcal{N}_I}$ and $\widehat{S}^z_{\mathcal{N}_I}$. 
The third sensitivity matrix $S^z_{I\mathcal{N}_I}\left(a_I\right)$ that is contained in the constraints of problem \labelcref{opt:identification problem distributed}, in contrast, is computed locally by subsystem $I$ in dependence on the optimization variable $a_I$.

Now that two different formulations of local identification problems have been presented, we briefly explain how a complete distributed ADI method is obtained from the local optimizations problem \labelcref{opt:identification problem local} or \labelcref{opt:identification problem distributed}, respectively, summarized as \Cref{algo:distributed adi}.
\begin{algorithm}[t]
	\caption{Distributed Attack Detection and Identification Based on Sparse Optimization}
	\label{algo:distributed adi}
	\begin{algorithmic}[1]% add [1] for line numbers
		\vskip2pt
		\Require
		\begin{minipage}[t]{\linewidth}%
			local dynamic model for each subsystem $I \in \mathcal{D}$ as in \labelcref{eq:distributed dynamic system discrete time},\\%
			\texttt{version} $\in \{1, 2\}$
		\end{minipage}%
		\State \texttt{detected} = \false, $a_I^\ast = 0$ for all $I$ \Comment{initialization}%
		\For{$I \in \mathcal{D}$}%
		\Comment{distributed attack detection}
		\State measure $z_I$, determine $\Delta z_I$%
		\If{$\|\Delta z_I\|_{\infty} > \tau_{\text{D}}$}%[local detection]
		\State \texttt{detected} = \true%
		\State \texttt{break}%
		\EndIf%
		\EndFor%
		\If{\texttt{detected}}%		
		\Comment{distributed attack identification}
		\For{$I \in \mathcal{D}$} 
		\If{\texttt{version} == 1}
		\State obtain coupling deviation $\Delta \widehat{z}_{\mathcal{N}_I}$ from neighbors
		\State solve local identification problem \labelcref{opt:identification problem local} to obtain $a_I^\ast$
		\Else
		\State obtain sensitivity information $\widehat{S}^a_{\mathcal{N}_I}, \widehat{S}^z_{\mathcal{N}_I}$ from neighbors
		\State solve local identification problem \labelcref{opt:identification problem distributed} to obtain $a_I^\ast$
		\EndIf
		\EndFor
		\EndIf\\
		\Return \texttt{detected}, $a_I^\ast$ for all $I$
	\end{algorithmic}
\end{algorithm}
The distributed detection scheme is based on monitoring the coupling variables and raises an alarm if an abnormal deviation $\Delta z_I > \tau_{\text{D}}$ is observed in any subsystem $I$. 
Then, the identification procedure is initiated and neighboring subsystems exchange the necessary information to set up the identification problem \labelcref{opt:identification problem local} or \labelcref{opt:identification problem distributed}, depending on which version is applied, and compute a solution to obtain a suspicion $a^{\ast}_I$ of the local attack.
If problem \labelcref{opt:identification problem distributed} is considered, the solution also suggests suspicions $a^{\ast}_{\mathcal{N}_I}$ and $\Delta \widehat{z}^{\ast}_{\mathcal{N}_{\mathcal{N}_I}}$ about the disturbing activities in the neighborhood.

Since the problem formulations in \labelcref{opt:identification problem local,opt:identification problem distributed} show some similarities to the global identification problem of our publication \cite{Braun2021Attack}, some of the theoretical considerations in \cite{Braun2021Attack} can be adopted with only minor changes.
E.g., an upper bound on the remainder term of the Taylor expansion can be obtained for the linearization error $R^{\text{lin}}_I$ in \labelcref{eq:taylor expansion for distributed adi}, when adapting the reasoning of \cite{Braun2021Attack} to the fact that here the expansion is only applied in the $\widehat{z}_{\mathcal{N}_I}$-argument but not the input.
The major difference between the identification problems for global versus distributed ADI is, however, that the constraints in problem \labelcref{opt:identification problem local,opt:identification problem distributed} are nonlinear, whereas a linear problem is considered in \cite{Braun2021Attack}.
As a consequence, the theoretical results from \cite{Candes2005Decoding} on relaxing the $\ell_0$-``norm'' cost term in compressed sensing problems by the $\ell_1$-norm are not applicable here since Candes and Tao restrict their considerations to linear constraints.
In fact, there is a body of research on nonlinear compressed sensing, e.g., \cite{Blumensath2013Compressed,Beck2013Sparsity}, the results of which can be useful to prove rigorous guarantees for the distributed ADI method presented in this section.
However, a precise elaboration of such proofs is out of scope for this paper and a promising direction for future work.

\section{Resilient Distributed MPC}
\label{sec:resilient dmpc}
While methods for attack identification are a very powerful tool to localize a priori unknown attacks and thus improve the resilience of control systems under malicious disturbances, they cannot prevent future attacks or reduce their impact.
On the other hand, robust control schemes can limit the impact of a perturbation by ensuring that no constraints are violated, but require information about the value range in which possible disturbances will lie, which is typically not available for unknown adversarial attacks.
We combine the advantages of both approaches by embedding the proposed ADI method into a DMPC setup, thus utilizing the identified insights about the attacker toward targeted robust DMPC.
To this end, we first describe an existing approach for robust DMPC in \Cref{subsec:contract based mpc}, and enhance it with \Cref{algo:distributed adi} to obtain an adaptively robust DMPC scheme in \Cref{subsec:adaptively robust} that computes robust control inputs against previously identified attacks in a distributed manner.

\subsection{Contract-Based Robust Distributed MPC}
\label{subsec:contract based mpc}
By \emph{robust} control, we refer to computing control inputs that ensure all constraints to a system with uncertain influences being met in all possible cases.
In \cite{Lucia2013Multi}, Lucia et al.\ introduce a multi-stage scheme for robust \emph{nonlinear MPC (NMPC)}, which considers discrete sets of scenarios and represents the possible evolution of the system state in a scenario tree like the one shown in \Cref{fig:scenario tree local}.
\def\unit{0.42}%
\def\dLevel{5*\unit}%
\def\dSibTwo{1.4*\unit}%
\def\dSibOne{3*\dSibTwo}%
\def\ellipseWidth{1.5*\unit}%
\def\eps{0.22*\unit}%
\def\epstwo{0.9*\unit}%
\def\epssource{0.5*\unit}%
\def\epslabel{1.6*\unit}%
\tikzstyle{level 1}=[level distance=\dLevel cm, sibling distance=\dSibOne cm]%
\tikzstyle{level 2}=[level distance=\dLevel cm, sibling distance=\dSibTwo cm]%
\newcommand{\scenarioTreeWithReachableSets}{%
\begin{tikzpicture}[grow = right, edge from parent/.style = {draw,-{Latex[length=0.3*\unit cm]}}, label distance = -0.42*\unit cm, every node/.style = {circle, fill=black, minimum width=\unit em, inner sep=\unit pt}, level distance = 3.1*\unit cm]%
%%% Draw reachable set around tree
\coordinate (v1) at (0, 0+\epssource) {};%
\coordinate (v2) at (\dLevel, \dSibOne+\epstwo+2.5*\eps) {};%
\coordinate (v3) at (2*\dLevel, \dSibOne+\dSibTwo+\epstwo+2*\eps) {};%
\coordinate (v4) at (4*\dLevel+2*\epstwo, \dSibOne+\dSibTwo+\epstwo+2*\eps) {};%
\coordinate (v5) at (4*\dLevel+2*\epstwo,-\dSibOne-\dSibTwo-\epstwo-\eps) {};%
\coordinate (v6) at (2*\dLevel,-\dSibOne-\dSibTwo-\epstwo-\eps) {};%
\coordinate (v7) at (\dLevel,-\dSibOne-\epstwo-2.5*\eps) {};%
\coordinate (v8) at (0, 0-\epssource-\eps) {};%
\fill [fill=gray!40]%
(v1) \foreach \i in {2,...,8}{ -- (v\i) } -- cycle;%
%
%%% Draw ellipses for approximations of reachable sets
\filldraw[fill=blue!50, draw=blue!50, line width=0.5mm] (\dLevel,0.05) ellipse (\ellipseWidth cm and \dSibOne cm + 4.4*\eps cm);%
\filldraw[fill=blue!50, draw=blue!50, line width=0.5mm] (2*\dLevel,0.1) ellipse (\ellipseWidth cm and \dSibOne cm + \dSibTwo cm + 4.7*\eps cm);%
\filldraw[fill=blue!50, draw=blue!50, line width=0.5mm] (3*\dLevel,0.1) ellipse (\ellipseWidth cm and \dSibOne cm + \dSibTwo cm + 4.7*\eps cm);%
\filldraw[fill=blue!50, draw=blue!50, line width=0.5mm] (4*\dLevel,0.1) ellipse (\ellipseWidth cm and \dSibOne cm + \dSibTwo cm + 4.7*\eps cm);%
\node[fill=none,label={270:{\textcolor{blue}{$\widetilde{\mathcal{X}}_I^{1,[0]}$}}}] at (1*\dLevel, -\dSibOne-\dSibTwo-\epslabel+2*\eps) {};%
\node[fill=none,label={270:{\textcolor{blue}{$\widetilde{\mathcal{X}}_I^{2,[0]}$}}}] at (2*\dLevel, -\dSibOne-\dSibTwo-\epslabel+2*\eps) {};%
\node[fill=none,label={270:{\textcolor{blue}{$\widetilde{\mathcal{X}}_I^{3,[0]}$}}}] at (3*\dLevel, -\dSibOne-\dSibTwo-\epslabel+2*\eps) {};%
\node[fill=none,label={270:{\textcolor{blue}{$\widetilde{\mathcal{X}}_I^{4,[0]}$}}}] at (4*\dLevel, -\dSibOne-\dSibTwo-\epslabel+2*\eps) {};%
%
%%% Draw scenario tree
\node[label={[label distance=-0.1*\unit cm]180:{\footnotesize$x_I^{0}$}}] {}%
    child {node[label={[label distance=-0.15*\unit cm]90:{\footnotesize$x_I^{1,s_7}$}}] {}%
        child {node[label=90:{\footnotesize$x_I^{2,s_9}$}] {}%
            child {node[label=90:{\footnotesize$x_I^{3,s_9}$}] {}%
            	child {node[label=90:{\footnotesize$x_I^{4,s_9}$}] {}}%
        	}%
                }%
        child {node[label=90:{\footnotesize$x_I^{2,s_8}$}] {}%
			child {node[label=90:{\footnotesize$x_I^{3,s_8}$}] {}%
				child {node[label=90:{\footnotesize$x_I^{4,s_8}$}] {}}%
			}%
		}%
        child {node[label=90:{\footnotesize$x_I^{2,s_7}$}] {}%
			child {node[label=90:{\footnotesize$x_I^{3,s_7}$}] {}%
				child {node[label=90:{\footnotesize$x_I^{4,s_7}$}] {}}%
			}%
		}%
    }%
    child {node[label=90:{\footnotesize$x_I^{1,s_4}$}] {}%
		child {node[label=90:{\footnotesize$x_I^{2,s_6}$}] {}%
			child {node[label=90:{\footnotesize$x_I^{3,s_6}$}] {}%
				child {node[label=90:{\footnotesize$x_I^{4,s_6}$}] {}}%
			}%
		}%
		child {node[label=90:{\footnotesize$x_I^{2,s_5}$}] {}%
			child {node[label=90:{\footnotesize$x_I^{3,s_5}$}] {}%
				child {node[label=90:{\footnotesize$x_I^{4,s_5}$}] {}}%
			}%
		}%
		child {node[label=90:{\footnotesize$x_I^{2,s_4}$}] {}%
			child {node[label=90:{\footnotesize$x_I^{3,s_4}$}] {}%
				child {node[label=90:{\footnotesize$x_I^{4,s_4}$}] {}}%
			}%
		}%
	}%
    child {node[label=90:{\footnotesize$x_I^{1,s_1}$}] {}%
		child {node[label=90:{\footnotesize$x_I^{2,s_3}$}] {}%
			child {node[label=90:{\footnotesize$x_I^{3,s_3}$}] {}%
				child {node[label=90:{\footnotesize$x_I^{4,s_3}$}] {}}%
			}%
		}%
		child {node[label=90:{\footnotesize$x_I^{2,s_2}$}] {}%
			child {node[label=90:{\footnotesize$x_I^{3,s_2}$}] {}%
				child {node[label=90:{\footnotesize$x_I^{4,s_2}$}] {}}%
			}%
		}%
		child {node[label=90:{\footnotesize$x_I^{2,s_1}$}] {}%
			child {node[label=90:{\footnotesize$x_I^{3,s_1}$}] {}%
				child {node[label=90:{\footnotesize$x_I^{4,s_1}$}] {}}%
			}%
		}%
	};%
\end{tikzpicture}%
}%

\begin{figure}[t]
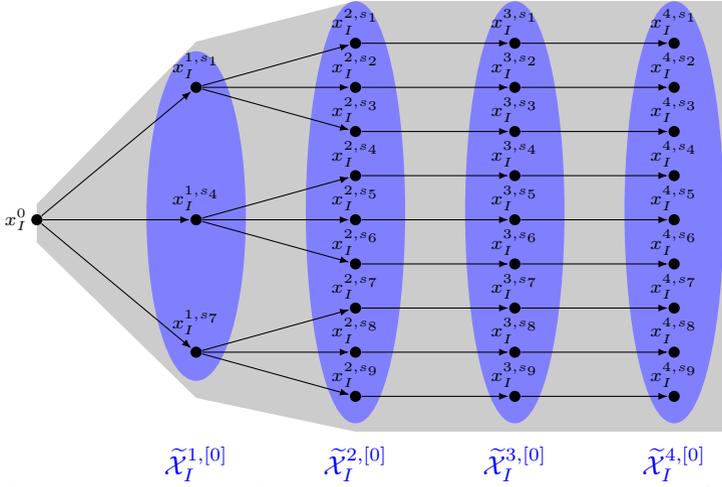
%
	\centering%
	\scenarioTreeWithReachableSets%
	\vspace{-0.3cm}%
	\caption{A scenario tree as in the multi-stage approach to robust MPC \cite{Lucia2013Multi}, here shown for time $k = 0$ and $N_{\text{p}}=4$, provides a natural and computationally efficient way to approximate the reachable sets $\mathcal{X}_I^{l, [k]}$ (indicated in gray) by discrete node sets $\widetilde{\mathcal{X}}_I^{l, [k]}$ (blue) explored by the tree.
	}%
	\label{fig:scenario tree local}%
\end{figure}%
In a distributed dynamic system, the neighbors' couplings $z_{\mathcal{N}_I}$ behave in an uncertain way to the eyes of subsystem $I$, and, therefore, robust MPC can also be used to design distributed MPC methods as long as each subsystem is provided with information about the range of possible neighboring coupling values.
In \cite{Lucia2015Contract}, this idea is implemented by Lucia et al.\ introducing so-called \emph{contracts}~$\mathcal{Z}_I$, which are corridors containing predicted reachable values of the coupling variables $z_I$ and are exchanged among neighbors. 
At time~$k$, the reachable state set $\mathcal{X}_I^{l+1, [k]}$ of all values that the local state $x_I^{l+1}$ may attain at time $l+1$ under all possible uncertainty realizations, is computed as
\begin{align*}
	\mathcal{X}_I^{l+1, [k]} \coloneqq &\left\{f_I\left(x_I^l, u_I^l + a_I^l, \widehat{z}^l_{\mathcal{N}_I}, w_I^l\right) : \right.\\
	&\left.\hphantom{\Big\{} x_I^l \in \mathcal{X}_I^{l, [k]}, a_I^l \in \mathcal{A}_I^{l, [k-1]}, \widehat{z}^l_{\mathcal{N}_I} \in \widehat{\mathcal{Z}}^{l, [k-1]}_{\mathcal{N}_I}, w_I^l \in \mathcal{W}_I^{l, [k-1]}\right\}
\end{align*}
with $\mathcal{X}_I^{k,[k]} \coloneqq \{x_I^k\}$.
From this, the contract $\mathcal{Z}_I^{l,[k]}$ for $z_I^l$ at time $k$ is derived as 
\begin{align*}
	\mathcal{Z}_I^{l, [k]} \coloneqq \left\{h_I\left(x_I^l\right): x_I^l \in \mathcal{X}_I^{l, [k]}\right\}.
\end{align*}
Similarly, contracts $\widehat{\mathcal{Z}}_I^{l,[k]}$ for the coupling coefficients $\widehat{z}_I^l$ are obtained using the dense coupling function $\widehat{\zeta}$.
These sets are computed locally at time $k$, provided that each subsystem knows attack and parameter uncertainty sets $\mathcal{A}_I^{l,[k-1]}$ and $\mathcal{W}_I^{l,[k-1]}$ and additionally receives its neighbors' contracts $\widehat{\mathcal{Z}}_I^{l,[k-1]}$.
If all these uncertainty sets are discrete or subsystem $I$ chooses finite subsets as sample scenarios, it can locally build a scenario tree as in \Cref{fig:scenario tree local}.
The tree contains one node $x_I^{l,s}$ for each time $l \in \{k, \dots, k + N_{\text{p}}\}$ with prediction horizon~$N_{\text{p}}$ and each scenario $s \in \Sigma_I^{[k-1]}$, where $\Sigma_I^{[k-1]}$ is the finite local index set of scenario indices $s$.
The local scenario trees allow to efficiently compute finite approximations $\widetilde{\mathcal{X}}_I^{l, [k]}$ of the reachable sets $\mathcal{X}_I^{l, [k]}$ as the set of tree nodes $x_I^{l,s}$ that are reached by subsystem $I$ at stage $l$ in any scenario $s \in \Sigma_I^{[k - 1]}$. This is indicated by blue shapes in \Cref{fig:scenario tree local} and explained in detail in \cite{Braun2020Hierarchicala}.
Corresponding approximated contracts $\widetilde{\mathcal{Z}}_I^{l, [k]}$ are obtained as 
\begin{align*}
	\widetilde{\mathcal{Z}}_I^{l,[k]} \coloneqq \left\{\widehat{\zeta}_I\left(x_I^{l,s}, u_I^{l,s}+a_I^{l,s}, \widehat{z}_{\mathcal{N}_I}^{l,s}, w_I^{l,s}\right): s\in \Sigma_I^{[k - 1]}\right\} \subseteq \widehat{\mathcal{Z}}_I^{l,[k]}
\end{align*}
and have been proven to work well in practice \cite{Braun2020Hierarchicala,Braun2021Adaptively}.
Considering every possible evolution of the uncertain system for the future time steps $k, \dots, k+N_{\text{p}}$ according to the finite scenario set $\Sigma_I^{[k - 1]}$, contract-based DMPC using multi-stage NMPC computes robust control inputs $u_I^k, \dots, u_I^{k+N_{\text{p}}-1}$ according to the following optimal control problem based on the work of Lucia et al.\ in \cite{Lucia2015Contract,Lucia2013Multi}
\begin{align}
	&\min_{x_I^{l,s},u_I^{l,s}}
	&&\sum_{s \in \Sigma_I^{[k - 1]}} \alpha_I^s
	\sum_{l = k}^{k + N_{\text{p}}-1} \ell_I\left(x_I^{l,s}, u_I^{l,s} + a_I^{l,s}, \widehat{z}_{\mathcal{N}_I}^{l,s}, w_I^{l,s}\right) 
	\notag\\
	&\hspace{2pt}\text{s.t.}
	&&x_I^{k,s} = x_I^{k}, \notag\\
	&&&x_I^{l+1,s} = f_I\left(x_I^{l,s}, u_I^{l,s} + a_I^{l,s}, \widehat{z}_{\mathcal{N}_I}^{l,s}, w_I^{l,s}\right),\notag\\
	&&&g_I\left(x_I^{l,s}, u_I^{l,s} + a_I^{l,s}, \widehat{z}_{\mathcal{N}_I}^{l,s}, w_I^{l,s}\right) \leq 0, \label{opt:ocp distributed under uncertainty relaxed}\\
	&&&x_I^{l+1,s} \in \mathbb{X}_I, u_I^{l,s} \in \mathbb{U}_I,\notag\\
	&&&x_I^{l,s} = x_I^{l,s'}
	~\Rightarrow~ u_I^{l,s} = u_I^{l,s'}, \notag\\
	&&&\min\left(\widetilde{\mathcal{Z}}_I^{l,[k-1]}\right)
	\leq 
	\widehat{\zeta}_I\left(x_I^{l,s}, u_I^{l,s} + a_I^{l,s}, \widehat{z}_{\mathcal{N}_I}^{l,s}, w_I^{l,s}\right)
	\leq 
	\max\left(\widetilde{\mathcal{Z}}_I^{l,[k-1]}\right),\notag\\
	&\hspace{2pt}\text{for all}
	&&s \in \Sigma_I^{[k - 1]}, s' \in \Sigma_I^{[k - 1]}, l \in \left\{k, \dots, k + N_{\text{p}} - 1\right\}.\notag
\end{align}
An optimal solution of problem \labelcref{opt:ocp distributed under uncertainty relaxed} provides a set of state trajectories starting at $x_I^k$ for all scenarios, behaving according to the local discrete-time dynamics as in \labelcref{eq:distributed dynamic system discrete time}, and taking only feasible states $x_I^{l+1,s} \in \mathbb{X}_I$. 
The optimal inputs are chosen to be feasible, to satisfy the constraints in \labelcref{eq:local constraints} in all scenarios $s \in \Sigma_I^{[k - 1]}$ and at all times $l$, and to minimize the local costs $\ell_I$ weighted over all scenarios with weights $\alpha_I^s \in \mathbb{R}_{\geq 0}$.
The problem formulation takes into account that future control inputs can be adapted when new measurements are available, while input values $u_I^{l,s}$, $u_I^{l, s'}$ that are applied to the same tree node have to coincide because a real-time controller cannot anticipate the future.
Finally, for consistency, we require each element $\widehat{z}_I^{l,s}$ of the updated contract $\widetilde{\mathcal{Z}}_I^{l,[k]}$ to be within the bounds of the previous contract $\widetilde{\mathcal{Z}}_I^{l,[k-1]}$.
For details on the purpose and the theoretical consequences of the last two groups of constraints we refer to the original works \cite{Lucia2015Contract,Lucia2013Multi} and our own work \cite{Braun2020Hierarchicala}.

\subsection{Adaptively Robust Distributed MPC}
\label{subsec:adaptively robust}
While we have explained in \Cref{subsec:contract based mpc} how updated contracts $\widetilde{\mathcal{Z}}_I^{l,[k]}$ are calculated at each time $k$ from a solution of problem \labelcref{opt:ocp distributed under uncertainty relaxed}, we have not yet commented on how to obtain similar scenario sets $\widetilde{\mathcal{A}}_I^{l,[k]}$ and $\widetilde{\mathcal{W}}_I^{l,[k]}$ for unknown attacks~$a_I^l$ and uncertain parameters $w_I^l$. 
For the latter, suitable samples are usually provided by forecasts, historical data, or technical properties of the system components.
For unknown attacks, however, it would be very restrictive to assume that appropriate scenario sets $\widetilde{\mathcal{A}}_I^{l,[k]}$ are provided. 
Choosing few random attacks as samples as in \cite{Braun2020Hierarchicala} cannot be expected to achieve satisfied constraints in all cases, while choosing a very large number of samples may cover the set $\mathbb{A}_I$ of possible attacks sufficiently well, but leads to computationally intractable problems since the size of the scenario tree grows exponentially in the number of scenarios.
To address this issue, we proposed a more general, adaptively robust MPC approach in \cite{Braun2021Adaptively} that utilizes available knowledge about the attackers gained from attack identification to design the sets $\widetilde{\mathcal{A}}_I^{l,[k]}$ and is repeated in this section.
Unlike in \cite{Braun2021Adaptively}, here the \emph{distributed} ADI approaches from \Cref{sec:dadi} are embedded in a DMPC setup, resulting in a fully distributed control framework that does not require any central instance.
The approach has already been described in \cite{Braun2022Resilient} and is presented here in further depth.

The method is designed for local attacks $a_I$ that follow a probability distribution with unknown, time-invariant expected value $\mu_I \in \mathbb{R}^{n_{u_I}}$ and standard deviation $\sigma_I \in \mathbb{R}^{n_{u_I}}_{\geq 0}$.
The basic idea is to repeatedly estimate these parameters at each time $k$ based on the solutions $a_I^{\ast, l}$ of the local attack identification problem at previous times $l \leq k$, and to adapt the uncertainty sets $\widetilde{\mathcal{A}}^{l, [k]}$ for possible attacks $a^l$ accordingly. 
More precisely, at time $k$ the mean $\mu_I^{[k]}$ and sample standard deviation $\sigma_I^{[k]}$ of all previously identified values $a_I^{\ast, l}$ given as 
\begin{align}
	\label{eq:update estimates for mean and std dev}
	\mu_I^{[k]} \coloneqq \frac{1}{k+1}\sum_{l=0}^{k} a_I^{\ast, l}  
	~~\text{ and }~~
	\sigma_I^{[k]} \coloneqq \left(\frac{1}{k} \sum_{l=0}^k \left(a_I^{\ast, l} - \mu_I^{[k]}\right)^2\right)^{\frac{1}{2}}
\end{align}
serve as estimates for $\mu_I$ and $\sigma_I$.
According to the local identification results until time $k$, the uncertainty of possible attacks $a_I^l$ for future time steps $l$ is represented by three scenarios for each component $(a_I^k)_i$ for $i \in \{1, \dots, n_{u_I}\}$
\begin{align}
	\label{eq:sampled attack set}
	\widetilde{\mathcal{A}}_I^{l,[k]} = \prod_{i \in I} \left\{\mu_i^{[k]}, \mu_i^{[k]} + \sigma_i^{[k]}, \mu_i^{[k]} - \sigma_i^{[k]}\right\}.
\end{align}%
The combination of contract-based robust DMPC from \Cref{subsec:contract based mpc} and the distributed ADI method from \Cref{sec:dadi} results in an adaptively robust distributed MPC method that is summarized in \Cref{algo:adaptively robust dmpc}.
\begin{algorithm}[h]
	\caption{Adaptively robust distributed MPC}
	\label{algo:adaptively robust dmpc}
	\begin{algorithmic}[1]% add [1] for line numbers
		\vskip2pt
		\Require 
		\begin{minipage}[t]{\linewidth}
			local dynamic model for each subsystem $I \in \mathcal{D}$,\\
			initial contracts $\widetilde{\mathcal{Z}}_I^{l,[0]}$ for all $I, l$, e.g., $\widetilde{\mathcal{Z}}_I^{l,[0]} = \{h_I(x_I^0)\}$, \\
			finite parameter scenario sets $\widetilde{\mathcal{W}}_I^{l,[k]}$ for all $l,k$
		\end{minipage}
		\State set $\widetilde{\mathcal{A}}_I^{l,[0]} \coloneqq \{\}$ for all $I,l$
		\For{time step $k$} 
		\For{$I \in \mathcal{D}$}
		\State build scenario tree by branching on $\widetilde{\mathcal{A}}_I^{l, [k-1]}$, $\widetilde{\mathcal{Z}}^{l, [k-1]}_{\mathcal{N}_I}$, and $\widetilde{\mathcal{W}}_I^{l, [k-1]}$
		\State solve problem \labelcref{opt:ocp distributed under uncertainty relaxed} to compute inputs $u_I^l$
		\State derive new contracts $\widetilde{\mathcal{Z}}_I^{l,[k]}$
		\Comment{update contracts}
		\State transmit $\widetilde{\mathcal{Z}}_I^{l,[k]}$ to neighbors
		\EndFor
		\State apply first control input $u^k = (u_I^k)_{I \in \mathcal{D}}$
		\For{$I \in \mathcal{D}$}
		\State solve problem \labelcref{opt:identification problem local} to obtain a suspicion $a_I^{\ast,k}$
		\Comment{local ADI}
		\State update estimates $\mu_I^{[k]}$, $\sigma_I^{[k]}$ as in \labelcref{eq:update estimates for mean and std dev}
		\State adapt uncertainty set $\widetilde{\mathcal{A}}_I^{l,[k]}$ as in \labelcref{eq:sampled attack set}
				\Comment{update attack scenarios}
		\EndFor		
		\EndFor			
	\end{algorithmic}
\end{algorithm}

We formulate \Cref{algo:adaptively robust dmpc} involving the local identification problem \labelcref{opt:identification problem local} and thus the first version of \Cref{algo:distributed adi} since this is what we apply in the numerical experiments presented in \Cref{sec:numerics}.
Clearly, \Cref{algo:adaptively robust dmpc} can also be defined based on the second version of \Cref{algo:distributed adi} solving problem \labelcref{opt:identification problem distributed}.
In this case, subsystem $I$ can additionally modify the transmitted contracts $\widetilde{\mathcal{Z}}_{\mathcal{N}_I}$ in such a way that the locally identified suspicions $a_{\mathcal{N}_I}^\ast$, $\Delta \widehat{z}_{\mathcal{N}_I}^\ast$ about neighboring attacks and coupling deviations are taken into account.
While this is not reasonable if the neighbors and thus their transmitted sensitivities $\widehat{S}^a_{\mathcal{N}_I}$ and $\widehat{S}^z_{\mathcal{N}_I}$ are generally deemed untrustworthy, it is useful if the communication channel to the neighbors is considered secure, but the neighbors themselves do not apply ADI and therefore do not adapt their contracts to attacks.

By enhancing distributed MPC with local attack identification in each subsystem, we obtain a distributed adaptively robust control framework, in which only locally available model knowledge and some information exchange among neighbors is involved.
Unlike the related method introduced in \cite{Braun2021Adaptively}, \Cref{algo:adaptively robust dmpc} requires no central authority and, in particular, no confidential model knowledge is published globally.
Such a procedure has the advantages that all local identification problems can be solved in parallel, that it can be employed even if the subsystems fail to agree on a central authority, and that no private model knowledge has to be shared with the entire network.
Furthermore, all distributed ADI approaches have in common that it is challenging to agree on system-wide countermeasures based on multiple, possibly contradictory local identification results. 
Our approach provides an answer to this issue as it transfers the insights from distributed ADI into \emph{local} countermeasures by adjusting the local control inputs in a suitable robust way.

\section{Dynamic Model for Microgrids Under Attack}
\label{sec:microgrid model}
Distributed microgrids that include local generation, demands, and often storage units, increase the security of supply within the microgrid area but create new challenges: Several optimal control tasks have to be addressed under the uncertainty of renewables and possibly even adversarial attacks, e.g., economic generator dispatch, efficient battery use, or optimal power import and export strategies to benefit from fluctuating energy prices \cite{Olivares2014Trends,Mohammed2019AC}.
Therefore, we aim to apply the resilient control framework proposed in \Cref{sec:resilient dmpc} to the task of microgrid control and derive a suitable dynamic model in this section.

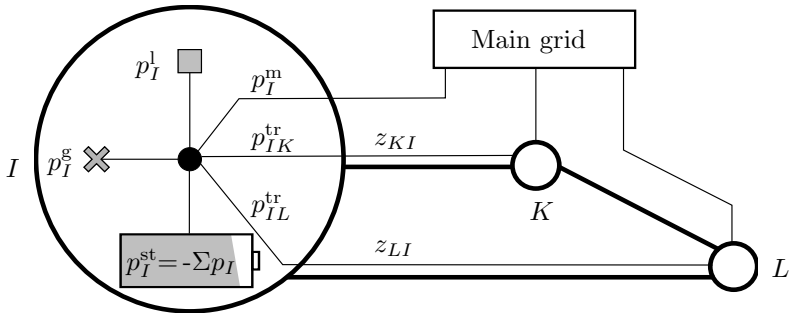
\begin{figure}[t]%
	\centering
	\begin{overpic}[width=0.8\linewidth]{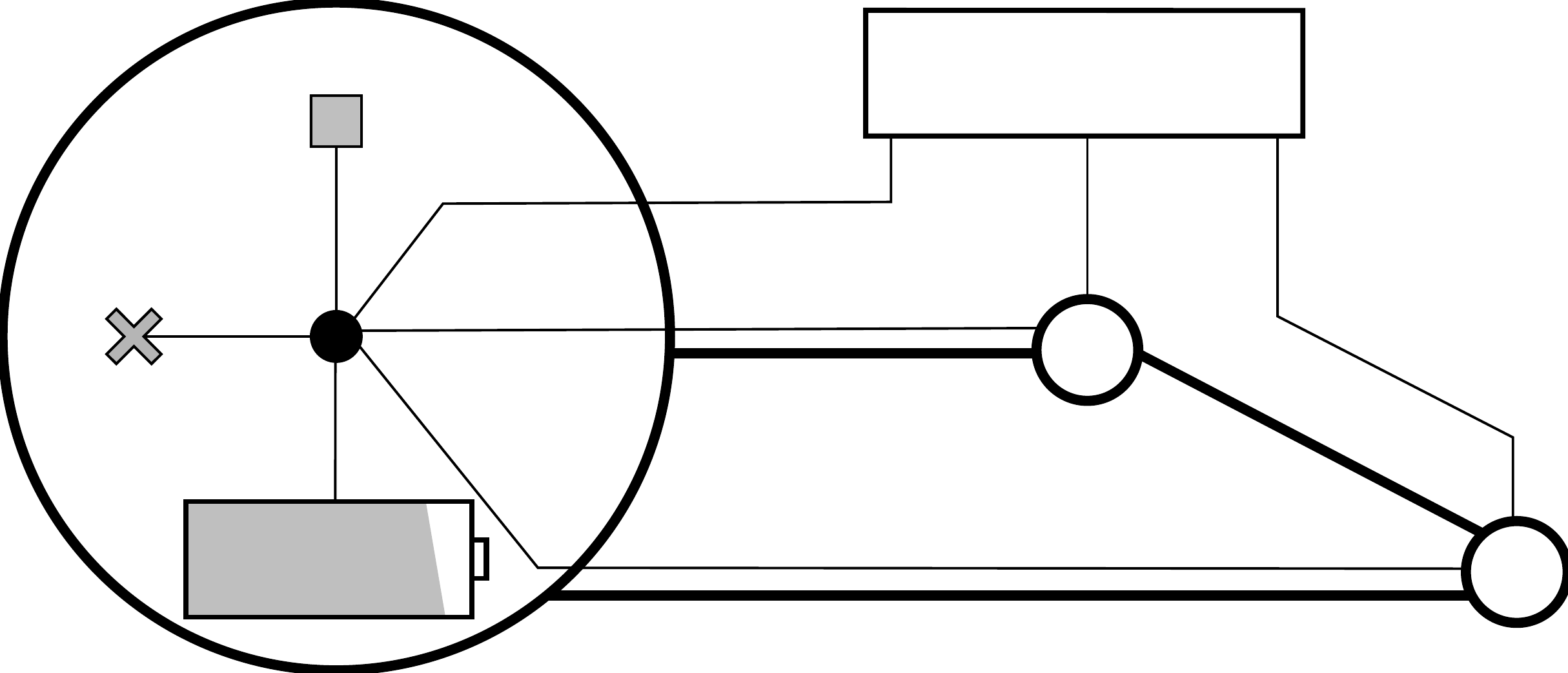}
		\put(12.7,6){$p^{\text{st}}_I\hskip-3pt=\hskip-1pt\text{-}\Sigma p_I$}
		\put(1.8,20){$p^{\text{g}}_I$}
		\put(14,33.5){$p^{\text{l}}_I$}
		\put(30,31.5){$p^{\text{m}}_I$}	
		\put(30,23.5){$p^{\text{tr}}_{IK}$}		
		\put(30,14.){$p^{\text{tr}}_{IL}$}		
		\put(-4,19){$I$}%i
		\put(102,5.){$L$}%j
		\put(68.5,13){$K$}%k
		\put(47,8.5){$z_{LI}$}%
		\put(47,23.5){$z_{KI}$}%
		\put(60.4,37){Main grid}
	\end{overpic}
	\caption{Schematic overview of the model for interconnected microgrids taken from \cite[Fig.\ 1]{Braun2022Resilient}, showing the local model components for microgrid $I$. Apart from internal states, each microgrid only requires knowledge of its neighboring couplings $(z_{LI})_{J\in \mathcal{N}_I}$. For power balance, storage units are used as a buffer.}%
	\label{fig:microgrid sample system}%
\end{figure}

The main characteristics of the model are nonlinear battery dynamics, physical coupling of neighboring microgrids through dispatchable power exchange, and the threat of possible attacks.
Each microgrid contains an aggregated load $p^{\text{l}}_I \leq 0$ and a set of dispatchable generation units that generate a total power output $p^{\text{g}}_I \geq 0$. 
How uncertain load and nondispatchable generation from renewable energy sources are modeled is discussed below.
As illustrated in \Cref{fig:microgrid sample system}, each microgrid is connected to the main grid, to or from which it can export or import power $p^{\text{m}}_I \in \mathbb{R}$. 
While power import is modeled by positive values $p^{\text{m}}_I > 0$, negative values $p^{\text{m}}_I < 0$ indicate power export to the main grid. 
In addition, power transfers are possible between two neighboring microgrids $I,L$ with $L \in \mathcal{N}_I$.
The power that microgrid $I$ provides to $L$ is denoted as $p^{\text{tr}}_{IL}$ and the resulting directed power flow from $I$ to $L$ is given as 
\begin{align*}
	p^{\text{flow}}_{IL}
	\coloneqq 
	p^{\text{tr}}_{IL} - p^{\text{tr}}_{LI}.
\end{align*} 
Finally, each microgrid has a storage unit that provides or consumes storage power $p^{\text{st}}_I \in \mathbb{R}$ and the state variable $s_I \in [0.0, 1.0]$ indicates its \emph{state of charge (SoC)}.
Power values $p^{\text{st}}_I > 0$ indicate discharging and $p^{\text{st}}_I < 0$ charging.
Unlike other works investigating economic dispatch problems in microgrid settings, for example Ananduta et al.\ in \cite{Ananduta2020Resilient}, we take into account that power cannot change instantaneously.
Instead, the dynamic evolution of~$p^{\text{g}}_I$, $p^{\text{m}}_I$, and $p^{\text{tr}}_{IL}$ is controlled by inputs $u^{\text{g}}_I$, $u^{\text{m}}_I$, and $u^{\text{tr}}_{IL}$ and behaves according to
\begin{align}%
	\dot{p}_I^{\text{g}} &= \frac{1}{T_I^{\text{g}}}\left(u^{\text{g}}_I + a^{\text{g}}_I - p^{\text{g}}_I\right), \label{eq:microgrid dynamics gen}\\ 
	\dot{p}_I^{\text{m}} &= \frac{1}{T_I^{\text{m}}}\left(u^{\text{m}}_I + a^{\text{m}}_I - p^{\text{m}}_I\right), \label{eq:microgrid dynamics main grid}\\
	\dot{p}^{\text{tr}}_{IL} &= \frac{1}{T^{\text{tr}}_{IL}}\left(u^{\text{tr}}_{IL} + a^{\text{tr}}_{IL} - p^{\text{tr}}_{IL}\right).\label{eq:microgrid dynamics transfer}
\end{align}%
The various delay parameters $T_I^{\text{g}}, T_I^{\text{m}}$, $T^{\text{tr}}_{IL} \in \mathbb{R}_{>0}$ depending on technical characteristics capture how quickly a change in the respective input affects the corresponding state. 
Compared to the generation delay $T_I^{\text{g}}$, typically smaller delay times $T_I^{\text{m}}$ and $T^{\text{tr}}_{IL}$ apply for power transfers with the main grid or neighboring microgrids.
In line with the generic description of distributed systems under attack introduced in \Cref{sec:problem}, we model attacks as additional, unknown inputs that impair the dynamic behavior of the microgrid systems as in \labelcref{eq:microgrid dynamics gen,eq:microgrid dynamics main grid,eq:microgrid dynamics transfer}.
In each microgrid $I \in \mathcal{D}$, we consider generator attacks $a^{\text{g}}_I \in \mathbb{R}$, grid attacks $a^{\text{m}}_I \in \mathbb{R}$ affecting the power exchange with the main grid, and transfer attacks $a^{\text{tr}}_{IL}  \in \mathbb{R}$ on power transfers to or from any neighbor $L \in \mathcal{N}_I$.
While the inputs are computed by the local controller in $I$, the attack values are unknown to the control system.
Thus, we deliberately make no difference in modeling attacks and renewable generation but consider both as uncertain influences resolved by the resilient control framework presented in \Cref{subsec:adaptively robust}.
Similarly, uncertain load can be considered an attack $a_I^{\text{l}}$ modifying the load $p^{\text{l}}_I = u_I^{\text{l}}$ that is modeled as a noncontrollable input with equal upper and lower bounds.

The storage is used as a buffer providing the required power reserves at all times and thus assuring that the power balance in microgrid $I$ is always satisfied, even when an attack occurs.  
Therefore, the storage power $p^{\text{st}}_I$ is a dependent variable according to
\begin{align*}
	p^{\text{st}}_I &= - p^{\text{g}}_I - p^{\text{m}}_I - p^{\text{l}}_I - \sum_{L \in \mathcal{N}_I} \left(p^{\text{tr}}_{LI} - p^{\text{tr}}_{IL}\right).
	%\label{eq:storage as buffer}
\end{align*}
It is important to distinguish that for microgrid $I$, the local state $p^{\text{tr}}_{IL}$ can be controlled via $u^{\text{tr}}_{IL}$ as in \labelcref{eq:microgrid dynamics transfer}, whereas the neighboring state $p^{\text{tr}}_{LI}$ is neither controllable nor is its dynamic behavior known by microgrid $I$. 
The physical interconnection of neighboring microgrids is instead modeled by a coupling variable $z_{LI} = p^{\text{tr}}_{LI}$ and is treated locally as an uncertain parameter as we discussed in detail in \Cref{subsec:contract based mpc}.
\Cref{fig:microgrid sample system} illustrates that the local knowledge is limited to local power variables and neighboring couplings.

According to the storage power $p^{\text{st}}_I$, the storage is charged or discharged and the resulting change in the SoC $s_I$ is modeled as
\begin{align*}
	\dot{s}_I &= b_I\left(s_I, p^{\text{st}}_I\right) %\label{eq:dynamics soc}
\end{align*}
with some function  $b_I: [0.0,1.0] \times \mathbb{R} \rightarrow \mathbb{R}$ modeling the battery dynamics.
While a linear approximation of this charging behavior is usually sufficient in the middle range of $[0.0, 1.0]$, it is not accurate for marginal values of the SoC which become extremely relevant in case of an attack.
Following the line of  \cite{Mathieu2016Controlling,Zhang2016Generalized}, the dynamics of the SoC are given as
\begin{align}
	\label{eq:battery dynamics i / Q}
	{\dot{s}_I} = - \frac{I_I^{\text{st}}}{Q_I^{\text{st}}},
\end{align}
with $Q_I^{\text{st}}$ denoting the maximum capacity of the battery and $I_I^{\text{st}}$ being the battery current.
Denoting the battery voltage by $U_I^{\text{st}}$, the storage power $p^{\text{st}}_I$ and the voltage $U_I^{\text{st}}$ are given as 
\begin{align}
	\label{eq:microgrid ocv and ohmic effect}
	p^{\text{st}}_I 
	=
	U_I^{\text{st}}I_I^{\text{st}} ~\text{ and }~
	U_I^{\text{st}} = U_I^{\text{OCV}}(s_I) + R_I^{\text{st}}I_I^{\text{st}}.
\end{align}
in line with \cite{Mathieu2016Controlling}.
The term $U_I^{\text{OCV}}$ denotes the \emph{open circuit voltage} (OCV), that depends on the SoC $s_I$, and the second summand determining $U_I^{\text{st}}$ models the ohmic effect with resistance $R_I^{\text{st}}$.
Rewriting \labelcref{eq:microgrid ocv and ohmic effect} results in the following relation for the storage power $p^{\text{st}}_I$:
\begin{align*}
	p^{\text{st}}_I = U_I^{\text{OCV}}(s_I)I_I^{\text{st}} +  R_I^{\text{st}}\left(I_I^{\text{st}}\right)^2.
\end{align*}
Solving this equation for $I_I^{\text{st}}$, the battery current $I_I^{\text{st}} = n_I\left(s_I, p^{\text{st}}_I\right)$ is obtained from $s_I$ and $p^{\text{st}}_I$ for some nonlinear function $n_I: [0.0,1.0] \times \mathbb{R} \rightarrow \mathbb{R}$.
Together with \labelcref{eq:battery dynamics i / Q}, this results in a nonlinear function
\begin{align*}
	b_I(s_I, p^{\text{st}}_I) 
	\coloneqq
	-\frac{n_I(s_I, p^{\text{st}}_I)}{Q_I^{\text{st}}}
\end{align*}
that describes the dynamic behavior of the battery.

It remains open to specify the open circuit voltage $U_I^{\text{OCV}}(s_I)$ using the model in \cite{Zhang2016Generalized}, that is accurate also for low and high SOCs: 
With parameters $\alpha_I, \beta_I, \gamma_I, \delta_I, \mu_I$, and~$\nu_I$ depending on the type of battery, the OCV is given by
\begin{align}
	\label{eq:microgrid ocv by zhang et al}
	U_I^{\text{OCV}}(s_I) 
	\coloneqq 
	\alpha_I + \beta_I (-\text{ln}(s_I))^{\mu_I} + \gamma_I s_I + \delta_I e^{\nu_I(s_I - 1)}.
\end{align}

Bringing all of the above together, we have characterized a distributed dynamic system of interconnected microgrids, which results in a model of the form as in \labelcref{eq:distributed dynamic system discrete time} when discretizing. Each microgrid is described by a local state 
\begin{align}
	\label{eq:microgrid model state}
	x_I = 
	\begin{pmatrix}
		s_I,
		&p^{\text{g}}_I,
		&p^{\text{m}}_I,
		&p^{\text{tr}}_I
	\end{pmatrix}^{\top}
	\in \mathbb{R}^{3+\lvert \mathcal{N}_I\rvert}
\end{align}
with $p^{\text{tr}}_I \coloneqq \left(p^{\text{tr}}_{IL}\right)_{L \in \mathcal{N}_I}$ and controlled by a local input
\begin{align}
	\label{eq:microgrid model input}
	u_I = 
	\begin{pmatrix}
		u^{\text{g}}_I,
		&u^{\text{m}}_I,	
		&u^{\text{tr}}_I
	\end{pmatrix}^{\top}
	\in \mathbb{R}^{2+\lvert\mathcal{N}_I\rvert},
\end{align}
that may be disturbed by an attack input
\begin{align}
	\label{eq:microgrid model attack}
	a_I = 
	\begin{pmatrix}
		a^{\text{g}}_I,
		&a^{\text{m}}_I,
		&a^{\text{tr}}_I
	\end{pmatrix}^{\top}
	\in \mathbb{R}^{2+\lvert\mathcal{N}_I\rvert}
\end{align}
with $u^{\text{tr}}_I \coloneqq \left(u^{\text{tr}}_{IL}\right)_{L \in \mathcal{N}_I}$ and $a^{\text{tr}}_I \coloneqq \left(a^{\text{tr}}_{IL}\right)_{L \in \mathcal{N}_I}$.
Power transfers to other microgrids physically couple neighboring microgrids to each other, which is modeled by local coupling variables
\begin{align}
	\label{eq:microgrid model coupling}
	z_I = 
	\begin{pmatrix}
		p^{\text{tr}}_{IL}
	\end{pmatrix}_{L \in \mathcal{N}_I}^{\top}
	~\text{ with }~
	z_{\mathcal{N}_I} 
	= 
	\begin{pmatrix}
		p^{\text{tr}}_{LI}
	\end{pmatrix}_{L \in \mathcal{N}_I}^{\top}.
\end{align}

Each microgrid $I \in \mathcal{D}$ is operated locally to meet the respective load $p^{\text{l}}_I$ at the lowest possible cost according to some objective function $J_I: \mathbb{R}_{>0} \rightarrow \mathbb{R}$, which specifies the costs incurred during some time window $[0, T]$ of length $T \in \mathbb{R}_{>0}$ and is defined as
\begin{align}%
	\label{eq:microgrid cost function}
	J_I(T) 
	\coloneqq 
	\int_0^{T} q_I\left(p^{\text{g}}_I, p^{\text{tr}}_I, p^{\text{st}}_I\right) + \ell_I\left(p^{\text{flow}}_I, p^{\text{m}}_I\right) \dx{t} + m_I\left(s_I\left(T\right)\right).
\end{align}%
It consists of quadratic stage costs $q_I$, piecewise linear stage costs $\ell_I$, and terminal costs $m_I$. 
The quadratic costs $q_I: \mathbb{R}_{\geq 0} \times \mathbb{R}^{n_{z_I}} \times \mathbb{R} \rightarrow \mathbb{R}_{\geq 0}$ with cost parameters $C^{\text{g}}_I, C^{\text{tr}}_I, C^{\text{st}}_I \in \mathbb{R}_{\geq 0}$ are given as 
\begin{align*}
	q_I\left(p^{\text{g}}_I, p^{\text{tr}}_I, p^{\text{st}}_I\right) 
	\coloneqq 
	C^{\text{g}}_I\left(p^{\text{g}}_I\right)^2 
	+ 
	\sum_{L \in \mathcal{N}_I} C^{\text{tr}}_I\left(p^{\text{tr}}_{IL}\right)^2
	+ 
	C^{\text{st}}_I\left(p^{\text{st}}_I\right)^2.
\end{align*}
They capture the per-unit costs of using the units for power generation, power transfers to neighbors, and the respective storage operations. 
In contrast, the piecewise linear costs $\ell_I$ model the economic profit or loss from selling or buying energy in trade with neighbors or the main grid.
Defining the positive and negative part functions 
\begin{align*}
	(v)_+
	\coloneqq 
	\left\{
	\begin{aligned}
		0 ~&\text{ if } v < 0, \\ 
		v ~&\text{ if } v \geq 0,
	\end{aligned}
	\right.
	~\text{ and }~
	(v)_-
	\coloneqq 
	\left\{
	\begin{aligned}
		v ~&\text{ if } v < 0, \\ 
		0 ~&\text{ if } v \geq 0,
	\end{aligned}
	\right.
\end{align*}
the piecewise linear cost function $\ell_I: \mathbb{R}^{n_{z_I}} \times \mathbb{R} \rightarrow \mathbb{R}$ is given as
\begin{align*}
	\ell_I\left(p^{\text{flow}}_I, p^{\text{m}}_I\right) 
	\coloneqq
	&\sum_{L \in \mathcal{N}_I} C^{\text{flow,ex}}_{LI}\left(p^{\text{flow}}_{LI}\right)_- 
	+\sum_{L \in \mathcal{N}_I} C^{\text{flow,im}}_{LI}\left(p^{\text{flow}}_{LI}\right)_+ \\
	&+ C^{\text{m,ex}}_I\left(p^{\text{m}}_I\right)_- + C^{\text{m,im}}_I\left(p^{\text{m}}_I\right)_+
\end{align*}
for each microgrid $I \in \mathcal{D}$, with local export and import per-unit prices $C^{\text{flow,ex}}_{LI}$, $C^{\text{flow,im}}_{LI}$, $C^{\text{m,ex}}_I$, $C^{\text{m,im}}_I \in \mathbb{R}_{\geq 0}$, which may fluctuate throughout the day.
In the numerical example in \Cref{sec:numerics}, we will consider import prices that are considerably higher than the export prices and thus focus on small producers, for which in practice it is often more profitable to generate power for their own demand than to buy electricity from the main grid, and for which power exports to the grid are only worthwhile at times of high demand.
The terminal costs $m_I: [0.0,1.0] \rightarrow \mathbb{R}_{\geq 0}$ account for degradation costs of the battery as
\begin{align*}
	m_I(s_I) 
	\coloneqq
	C^{\text{dis}}_I\left(s_I(0) - s_I(T)\right)_ + Q_I^{\text{st}}.
\end{align*}
If the state of charge $s_I(T)$ at the end of the considered horizon is smaller than $s_I(0)$ at the beginning, each unit of power discharge is penalized by some cost $C^{\text{dis}}_I \in \mathbb{R}_{\geq 0}$.

\section{Numerical Experiments with Microgrids Under Attack}
\label{sec:numerics}
In this section, we present a numerical case study to analyze the performance of adaptively robust DMPC from \Cref{sec:resilient dmpc} in the context of interconnected microgrids under attack using the model from \Cref{sec:microgrid model}.
In contrast to our earlier work \cite{Braun2021Adaptively}, we apply \emph{distributed} ADI based on the local identification problem \labelcref{opt:identification problem local}. 
In the experiments, we address the question of how to achieve an economic operation of microgrids at minimum costs despite uncertainties.
Whether these emerge in form of disturbances with rather small impact, fluctuating generation from renewables, or malicious attacks; all represent critical yet all the more relevant threats to energy supply.

To this end, we consider three microgrids \subsys{1}, \subsys{2}, and \subsys{3} with renewable generation that are each connected to the main grid and the other two microgrids as in \Cref{fig:microgrid sample system}.
The initial values and bounds for all variables of the microgrid model are given in \Cref{mpc:tab:microgrid bounds} and the parameters are chosen as in \Cref{mpc:tab:microgrid parameter costs}, using those for lithium-titanate ($\text{Li}_4\text{Ti}_5\text{O}_{12}$) batteries from \cite{Zhang2016Generalized} in \labelcref{eq:microgrid ocv by zhang et al}.

\begin{table}[b]%
	\centering%
	\caption{This table lists lower and upper bounds as well as initial values at time $t = 0$ for all state and input variables of the microgrid model. For the state of charge, three distinct initial values $s_I(0)$ for the three microgrids \subsys{1}, \subsys{2}, and \subsys{3} are given. In all other cases, the indicated values apply for all subsystems.}%
	\label{mpc:tab:microgrid bounds}%
	\begin{tabular}{c|c|c|c|c}%
		{Variable} & {Lower Bound} & {Upper Bound} & {Initial Value} & {Unit}\\\hline%
		$s_I$ & 0.0 & 0.1 & 0.9, 0.5, 0.6 & - \\%
		$p^{\text{g}}_I$ & 0.0 & 1000.0 & 0.0 & \sinew{\kilo\watt} \\%
		$p^{\text{m}}_I$ & -1000.0 & 2000.0 & 0.0 & \sinew{\kilo\watt} \\%
		$p^{\text{tr}}_{IL}$ & -100.0 & 100.0 & 0.0 & \sinew{\kilo\watt} \\%
		%	\hline%
		$u^{\text{g}}_I$ & 0.0 & 1000.0 & - & \sinew{\kilo\watt} \\%
		$u^{\text{m}}_I$ & -1000.0 & 2000.0 & - & \sinew{\kilo\watt} \\%
		$u^{\text{tr}}_{IL}$ & -100.0 & 100.0 & - & \sinew{\kilo\watt}%
	\end{tabular}%
\end{table}%

For a timespan of two days, robust NMPC is applied locally with step size $\Delta t = 0.25\sinew{\hour}$ by each microgrid.
At time $t \in [0.0,48.0]\sinew{\hour}$, the local cost function $J_I$ in \labelcref{eq:microgrid cost function} takes into account the upcoming time window $[t, t + N_{\text{p}}]$ with prediction horizon $N_{\text{p}} = 6.0\sinew{\hour}$ and uses the cost parameters from \Cref{mpc:tab:microgrid parameter costs}. 
The values $C^{\text{m,im}}_I$ and $C^{\text{m,ex}}_I$, that describe the cost or revenue of power imports from or exports to the main grid, vary in the course of the day. 
In our example, we focus on microgrids that represent small local prosumers and use the following fictitious values for all microgrids, which are based on real prices on the German electricity market in 2021 \cite{Bundesnetzagentur2021Smard} and reflect typical market fluctuations with rising prices in the morning and evening hours:
\begin{align*}
	C^{\text{m,im}}_I(t) &= \left\{
	\begin{aligned}%
		275 ~&\text{ if } (t \,\text{mod} \,24\sinew{\hour})\in [15, 20)\sinew{\hour},\\	
		200 ~&\text{ if } (t \,\text{mod} \,24\sinew{\hour})\in [6, 9)\cup [20, 22)\sinew{\hour},\\		
		150 ~&\text{ if } (t \,\text{mod} \,24\sinew{\hour})\in [9, 15)\cup [22, 24)\sinew{\hour},\\		
		100 ~&\text{ otherwise},
	\end{aligned}\right.\\
	C^{\text{m,ex}}_I(t) &= \left\{
	\begin{aligned}
		15 ~&\text{ if } (t \,\text{mod} \,24\sinew{\hour})\in [15, 20)\sinew{\hour},\\	
		10 ~&\text{ if } (t \,\text{mod} \,24\sinew{\hour})\in [6, 9)\cup [20, 22)\sinew{\hour},\\
		0 ~&\text{ otherwise}.
	\end{aligned}\right.
\end{align*}
Here, $\text{mod}$ is the modulo operator and $(t\,\text{mod}\,24\sinew{\hour})$ denotes the time of day.

\begin{table}[t!]%
	\centering%
	\caption{This table lists all model and cost parameters that are used in the numerical experiments presented in this section. All values apply to all subsystems $I \in \{\subsys{1},\subsys{2},\subsys{3}\}$, except for $Q_I^{\text{st}}$, $R_I^{\text{st}}$, and $C^{\text{g}}_I$, where individual values for the respective subsystems are specified.}%
	\label{mpc:tab:microgrid parameter costs}%
	\subfloat[Model Parameters]{%
		\begin{tabular}{c|c|c}%
			Param. & Value & Unit \\\hline
			$p^{\text{l}}_I$ & -2.0 & \si{\kilo\watt} \\%
			$T^{\text{g}}_I$ & 0.1  & \si{\hour} \\%
			$T^{\text{m}}_I$ &0.001 & \si{\hour}\\%
			$T^{\text{tr}}_{IL}$ &0.001 & \si{\hour} \\%		
			$Q_I^{\text{st}}$ & 100, 200, 100 &\si{\kilo\amperehour} \\%
			$R_I^{\text{st}}$ & 1.5, 2.0, 3.0 &\si{\milli\ohm}
		\end{tabular}%
	}%
	\hfill%
	\subfloat[OCV Parameters]{%
		\begin{tabular}{c|c|c}%
			Param. & Value & Unit \\\hline
			$\alpha_I$ & 2.23 &\si{\volt} \\%
			$\beta_I$ &-0.001 &\si{\volt}\\%
			$\gamma_I$ &-0.35 &\si{\volt}\\%
			$\delta_I$ & 0.6851 &\si{\volt}\\%
			$\mu_I$ &3.0 & - \\%
			$\nu_I$ & 1.6 & -%
		\end{tabular}%
	}%
	\hfill%
	\subfloat[Cost Parameters]{%
		\begin{tabular}{c|c}%
			{Param.} & {Value}%
			\\\hline
			$C^{\text{g}}_I$ & 0.2, 3.0, 2.0\\
			$C^{\text{tr}}_I$ & 4.0\\
			$C^{\text{st}}_I$ & 1.0 \\
			$C^{\text{dis}}_I$ & 2000 \\
			$C^{\text{flow,im}}_{IL}$	& 4.0 \\%
			$C^{\text{flow,ex}}_{IL}$	& 0.04 %
		\end{tabular}%
	}%
\end{table}%
To achieve a resilient operation, the system is controlled using the adaptively robust distributed NMPC scheme from \Cref{subsec:adaptively robust}. 
Based on the local control problem \labelcref{opt:ocp distributed under uncertainty relaxed}, at each sampling time $k$ every microgrid computes contracts $\widetilde{\mathcal{X}}_I^{l,[k]}$ to confine the behavior of its future coupling values $z_I^{l}$ for $l \in \{k, \dots, k + N_{\text{p}} - 1\}$ and shares them with its neighbors. 
In contrast to the experiments in \cite{Braun2021Adaptively}, which involve a centralized ADI method, each microgrid consults \emph{locally} identified solutions $a_I^{\ast,k}$ of problem \labelcref{opt:identification problem local} to update its estimates $\mu_I^{[k]}$ and $\sigma_I^{[k]}$ of the expected value and standard deviation of the unknown random attack $a_I$ as in \labelcref{eq:update estimates for mean and std dev}.
In our numerical experiments, the nonlinear identification problem \labelcref{opt:identification problem local} is solved to an accuracy of $\varepsilon_I = 10^{-3}$ using the interior-point solver Ipopt \cite{Wachter2006Implementation}.
The states $x_I$ are assumed to be only \emph{partially} observable with linear output function $c_I: \mathbb{X}_I \rightarrow \mathbb{Y}_I$ that is defined as
\begin{align*}
	c_I\left(x_I\right)
	\coloneqq
	\text{diag}(1, 1, 1, 0, 0)
	x_I.
\end{align*}
This means that for each microgrid I, the outputs $y_I = (s_I, p^{\text{g}}_I, p^{\text{m}}_I)^{\top}$ are considered by the local identification process, but not the transfer variables $p^{\text{tr}}_{IL}$ for all $L \in \mathcal{N}_I$.
Based on the suspected attacks $a_I^{\ast,k}$ and the derived estimates $\mu_I^{[k]}$ and $\sigma_I^{[k]}$, the uncertainty sets $\widetilde{\mathcal{A}}_I^{l,[k]}$ are approximated as in \labelcref{eq:sampled attack set}. 
The local control problem \labelcref{opt:ocp distributed under uncertainty relaxed} is repeatedly adapted to new contracts and identification results that become available.
As a consequence, the inputs $u_I^{l}$ computed at time $k+1$ for $l \in \{k + 1, \dots, k + N_{\text{p}}\}$ are robust toward deviations in neighboring couplings within $\widetilde{\mathcal{Z}}_{\mathcal{N}_I}^{l,[k]}$ and identified attacks in $\widetilde{\mathcal{A}}_I^{l,[k]}$. 

\begin{figure}[t!]%
	\hspace{0.9cm}%
	\begin{minipage}[c]{\linewidth}%
		\begin{subfigure}{0.43\linewidth}%
			\begin{overpic}[width=\linewidth]{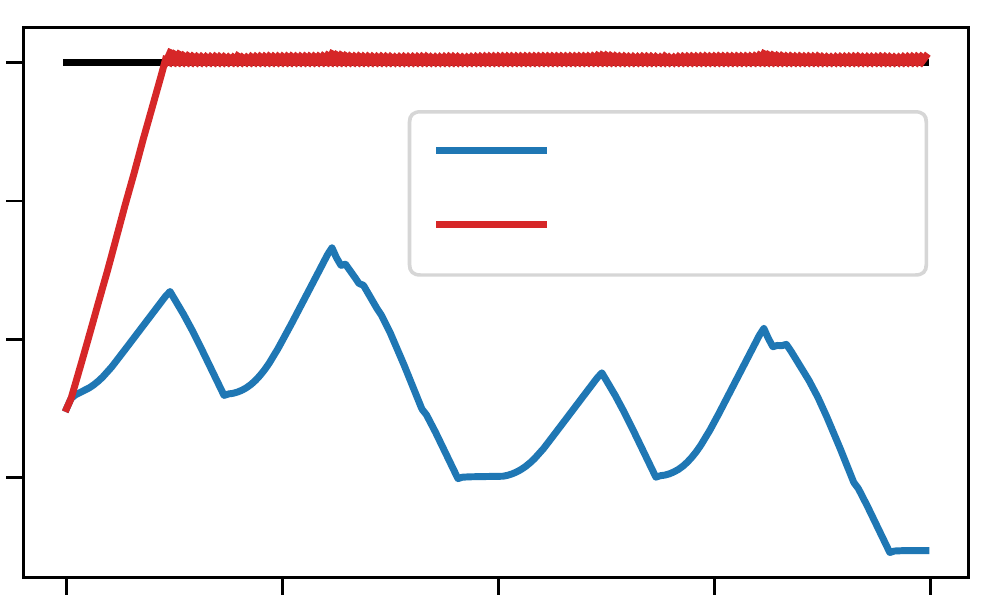}%
				% Legend entries
				\put(56.2, 44.4){\scriptsize $s_{\subsys{1}}$\,robust}%
				\put(56.2, 37.0){\scriptsize $s_{\subsys{1}}$\,non-robust}%
				% y-axis
				\put(-16, 21.4){\scriptsize \rotatebox{90}{SoC in \%}}%
				\put(-7.9, 53.2){\footnotesize 100}%		
				\put(-7.9, 39.2){\footnotesize \hphantom{1}96}%	
				\put(-7.9, 25.2){\footnotesize \hphantom{1}92}%				
				\put(-7.9, 11.2){\footnotesize \hphantom{1}88}%		
			\end{overpic}%
			\caption{State of Charge}%
			\label{fig:microgrid attack generation robust vs nonrobust trajectories:subfig:soc}%
		\end{subfigure}%
		\begin{subfigure}{0.43\linewidth}%
			\begin{overpic}[width=\linewidth]{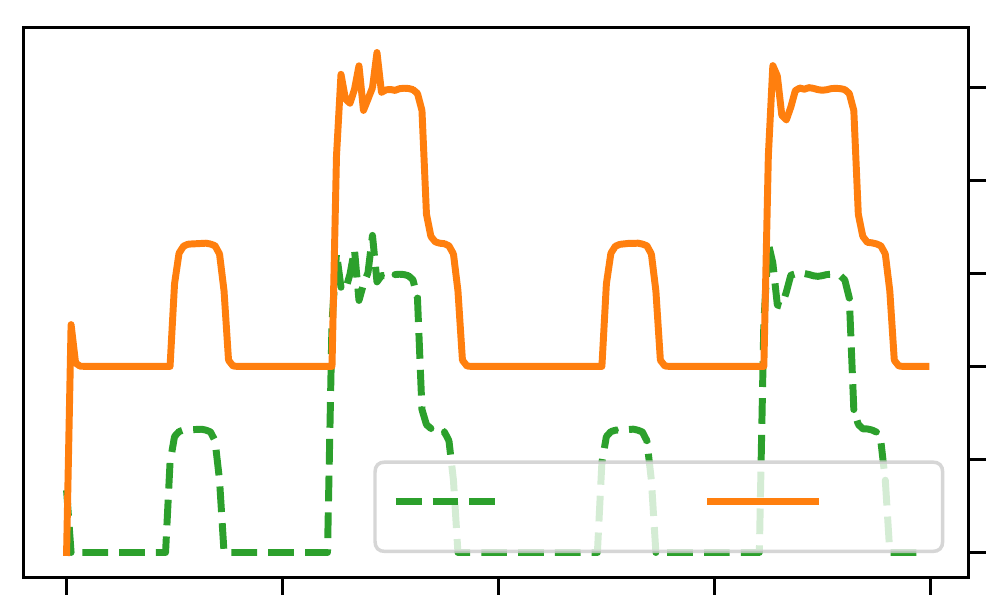}%
				% y-axis labels power generation	
				\put(111, 54){\scriptsize \rotatebox{-90}{Generation in \si{\kilo\watt}}}%	
				\put(100.5, 50.8){\footnotesize 25}%
				\put(100.5, 41.46){\footnotesize 20}%		    
				\put(100.5, 32.12){\footnotesize 15}%		    
				\put(100.5, 22.78){\footnotesize 10}%		    
				\put(100.5, 13.44){\footnotesize 5\hphantom{0}}%
				\put(100.5, 4.1){\footnotesize 0\hphantom{0}}%
				% Legend entries
				\put(51.8, 8.4){\scriptsize $u^{\text{g}}_{\subsys{1}}$}%
				\put(84., 8.4){\scriptsize $p^{\text{g}}_{\subsys{1}}$}%
			\end{overpic}%
			\caption{Power Generation}%
			\label{fig:microgrid attack generation robust vs nonrobust trajectories:subfig:generation}%
		\end{subfigure}%
		\newline%
		\begin{subfigure}{0.43\linewidth}%	
			\begin{overpic}[width=\linewidth]{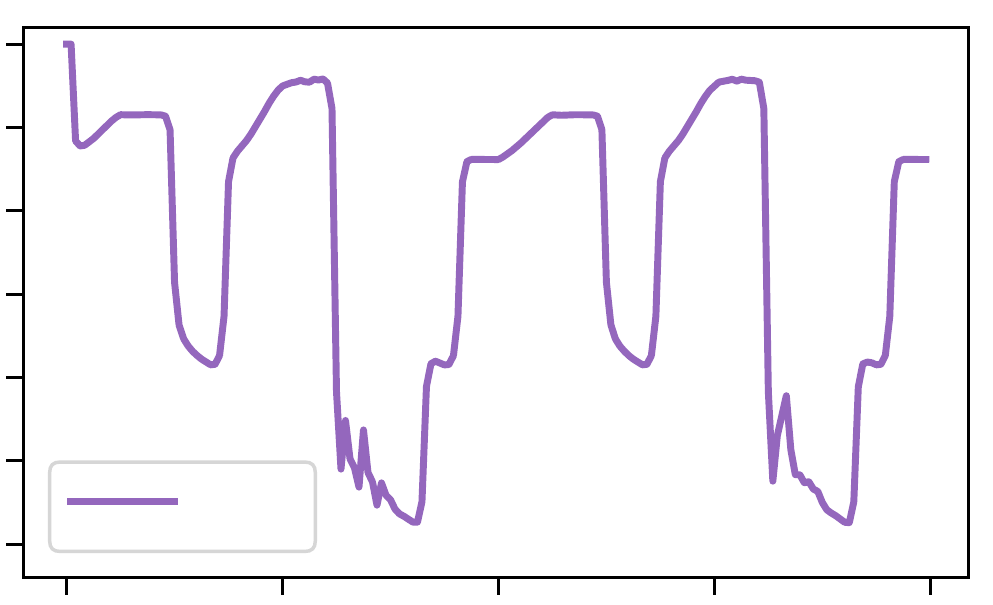}%
				% x-axis description and labels		
				\put(35, -9.4){\scriptsize Time in hours}%
				\put(3.4, -2.9){\footnotesize 0.0}%
				\put(24, -2.9){\footnotesize 12.0}%
				\put(45.8, -2.9){\footnotesize 24.0}%
				\put(67.5, -2.9){\footnotesize 36.0}%
				\put(89.3, -2.9){\footnotesize 48.0}%
				% y-axis labels
				\put(-16, -4){\scriptsize \rotatebox{90}{Imports / exports in \si{\kilo\watt}}}%			
				\put(-7.2, 55){\footnotesize \hphantom{-1}0}%
				\put(-7.2, 46.56){\footnotesize \hphantom{1}-5}%
				\put(-7.2, 38.13){\footnotesize -10}%
				\put(-7.2, 29.7){\footnotesize -15}%
				\put(-7.2, 21.26){\footnotesize -20}%
				\put(-7.2, 12.83){\footnotesize -25}%		
				\put(-7.2, 4.4){\footnotesize -30}%
				% Legend entries
				\put(21, 8.5){\scriptsize $p^{\text{m}}_{\subsys{1}}$}%
			\end{overpic}%
			\vspace{0.5cm}%
			\caption{Power exchange with main grid}%
		\end{subfigure}%
		\begin{subfigure}{0.43\linewidth}%
			\begin{overpic}[width=\linewidth]{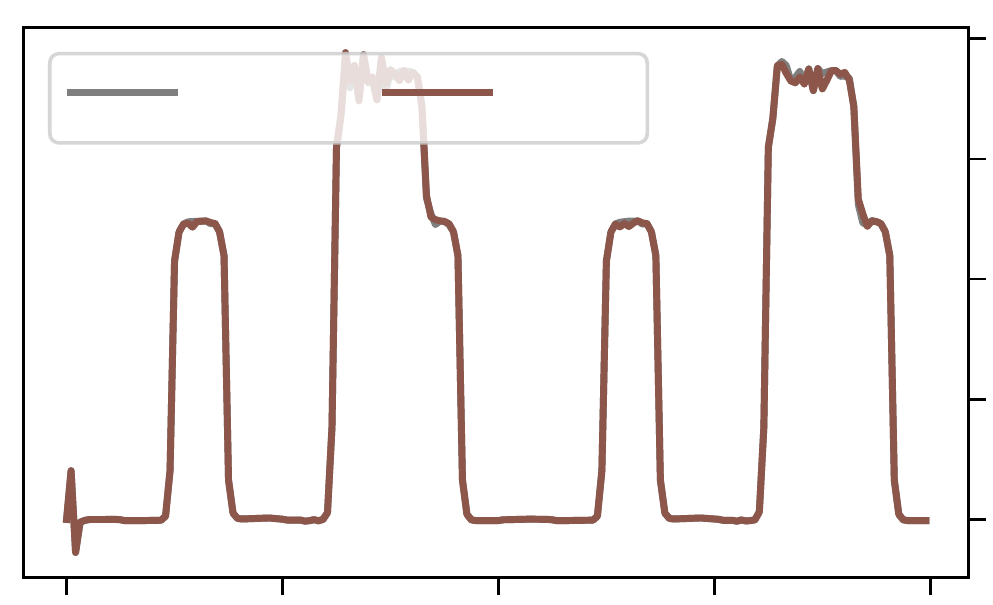}%
				% x-axis description and labels		
				\put(35, -9.4){\scriptsize Time in hours}%
				\put(3.4, -2.9){\footnotesize 0.0}%
				\put(24, -2.9){\footnotesize 12.0}%
				\put(45.8, -2.9){\footnotesize 24.0}%
				\put(67.5, -2.9){\footnotesize 36.0}%
				\put(89.2, -2.9){\footnotesize 48.0}%
				% y-axis
				\put(111, 51){\scriptsize\rotatebox{-90}{Transfers in \si{\kilo\watt}}}%
				% y-axis labels power transfer
				\put(100.5, 55.3){\footnotesize 2.0}%				
				\put(100.5, 43.25){\footnotesize 1.5}%				
				\put(100.5, 31.2){\footnotesize 1.0}%
				\put(100.5, 19.15){\footnotesize 0.5}%			
				\put(100.5, 7.1){\footnotesize 0.0}%			
				% Legend entries	
				\put(19.5, 49.5){\scriptsize $p^{\text{tr}}_{\subsys{1},\subsys{2}}$}%
				\put(51, 49.5){\scriptsize $p^{\text{tr}}_{\subsys{1},\subsys{3}}$}%
			\end{overpic}%
			\vspace{0.5cm}%
			\caption{Power exchange with neighbors}%
		\end{subfigure}%
	\end{minipage}%
	\caption{Selected state and input trajectories for microgrid \subsys{1}, showing all powers in \si{\kilo\watt}. The microgrid is exposed to a generator attack, causing the generation $p^{\text{g}}_{\subsys{1}}$ to be considerably larger than planned by $u^{\text{g}}_{\subsys{1}}$. The different SoC trajectories, computed by adaptively robust versus nonrobust NMPC, show the benefit of the proposed resilient control framework.% The figure is similar to \cite[Fig.\ 2]{Braun2022Resilient}.
	}%
	\label{fig:microgrid attack generation robust vs nonrobust trajectories}%
\end{figure}
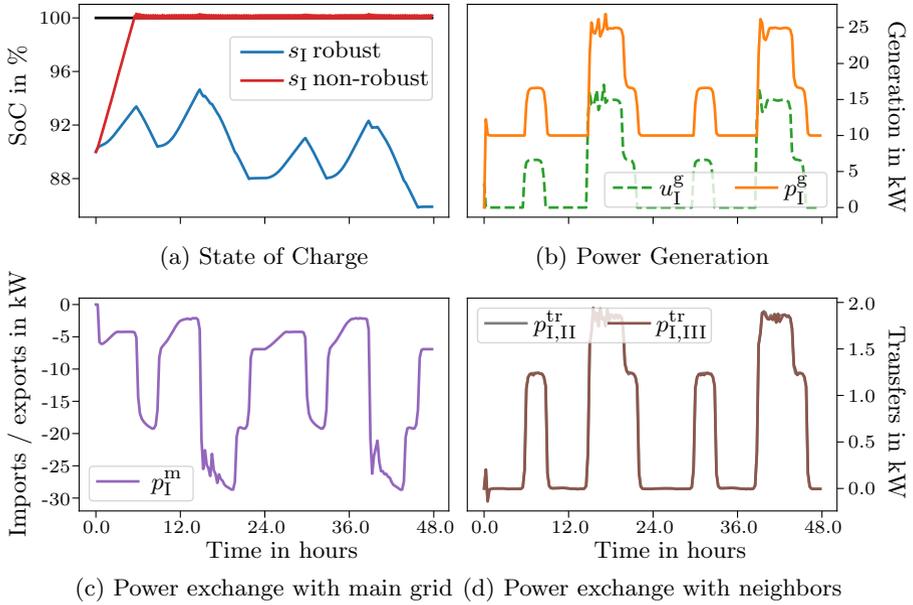
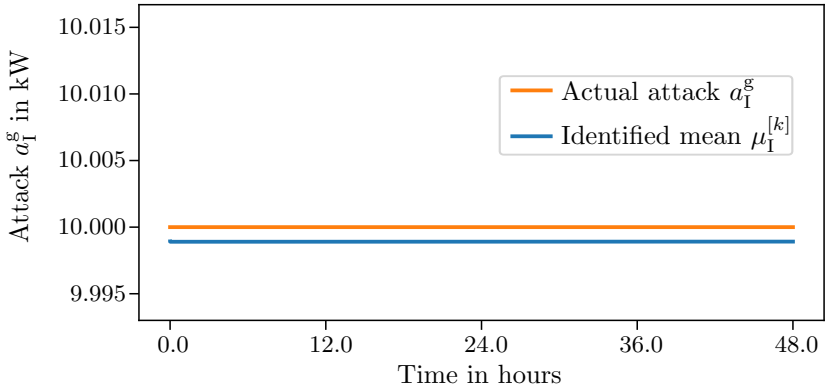
\begin{figure}[t!]%
	\hspace{1.7cm}%
	\begin{overpic}[width=0.8\linewidth]{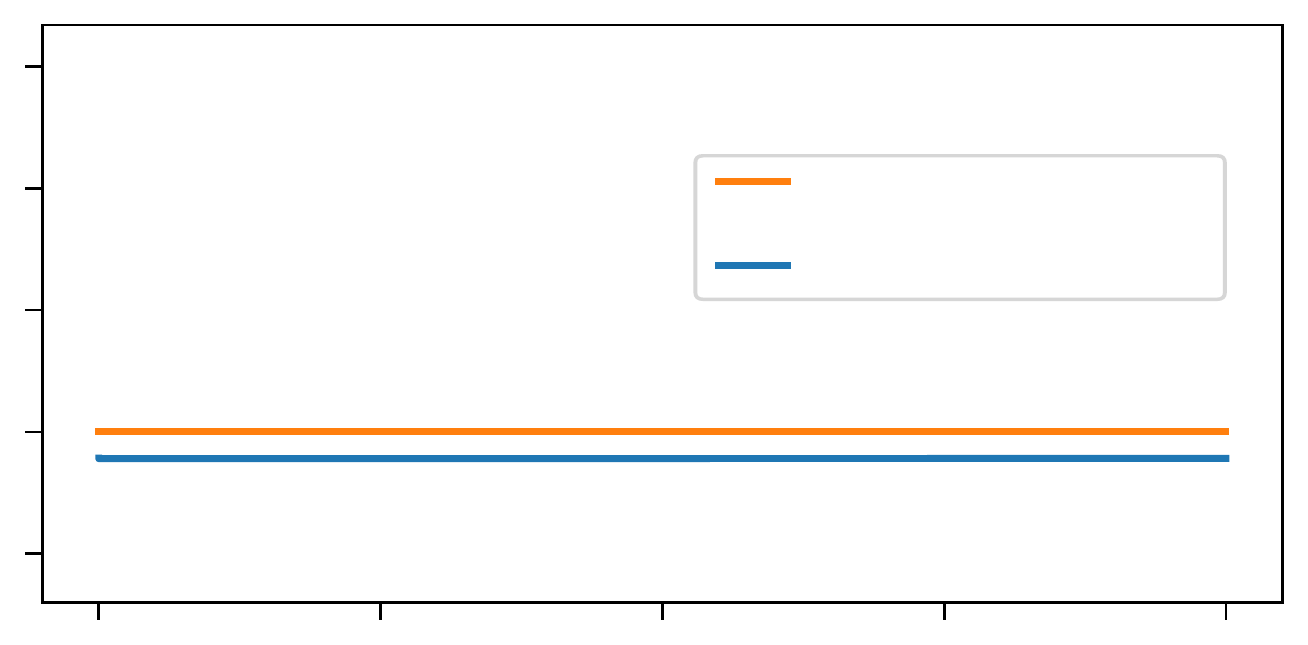}%
		\put(61.7, 34.1){Actual attack $a^{\text{g}}_{\subsys{1}}$}%
		\put(61.7, 27.9){Identified mean $\mu^{[k]}_{\subsys{1}}$}%
		\put(39, -5.5){Time in hours}%
		\put(5.8, -1.2){\small 0.0}%
		\put(26.8, -1.2){\small 12.0}%
		\put(48.3, -1.2){\small 24.0}%
		\put(69.8, -1.2){\small 36.0}%
		\put(91.3, -1.2){\small 48.0}%	
		\put(-15, 14.1){\rotatebox{90}{Attack $a^{\text{g}}_{\subsys{1}}$ in \si{\kilo\watt}}}%
		\put(-8.0, 43.3){\small 10.015}%
		\put(-8.0, 33.975){\small 10.010}%		
		\put(-8.0, 24.65){\small 10.005}%
		\put(-8.0, 15.325){\small 10.000}%
		\put(-8.0, 6.){\small \hphantom{1}9.995}%
	\end{overpic}%
	\vspace{0.8cm}%
	\caption[Comparison of the real and the identified attack computed by distributed ADI in a microgrid example]{Actual attack value $a^{\text{g}}_{\subsys{1}}$ and average identified value $\mu^{[k]}_{\subsys{1}}$ in the first attack scenario examined, in which only dispatchable generation units are in use and microgrid~\subsys{1} is exposed to a generator attack.}%
	\label{fig:microgrid adi results generation attack}%
\end{figure}

%
We examine the behavior of the system, controlled with \Cref{algo:adaptively robust dmpc}, in two attack scenarios. 
For comparison, we repeat each experiment with nonrobust DMPC, where neither contracts are exchanged nor attack identification is considered.
First, we assume that all generation units are dispatchable and a constant attack $a^{\text{g}}_{\subsys{1}} = 10.0\sinew{\kilo\watt}$ disrupts the generator dynamics in microgrid \subsys{1} according to \labelcref{eq:microgrid dynamics gen}. 
The attacker is active over the entire time window $[0.0,48.0]\sinew{\hour}$ and causes a severe deviation of the generated power $p^{\text{g}}_{\subsys{1}}$ in microgrid \subsys{1} from the control input $u^{\text{g}}_{\subsys{1}}$ as \Cref{fig:microgrid attack generation robust vs nonrobust trajectories} reveals. 
The distributed ADI method based on the local identification problem \labelcref{opt:identification problem local} successfully identifies the unknown attack input with very high precision in every time step as pointed out by \Cref{fig:microgrid adi results generation attack}, which shows the mean of the suspected attack values $a^{\text{g},\ast}_{\subsys{1}} \approx 9.9989\sinew{\kilo\watt}$ at all times.
This allows the local robust NMPC scheme to adapts its prediction very accurately and adjust the control inputs accordingly. 
As a result, the microgrid takes advantage of the additional power generation by charging the battery and exporting the power to the main grid during times with high profit. 
In the solution computed with nonrobust NMPC, on the contrary, the battery reaches and violates its maximum state of charge of $1.0$ after about $5.0\sinew{\hour}$ as the red SoC trajectory in \Cref{fig:microgrid attack generation robust vs nonrobust trajectories:subfig:soc} reveals. 
Due to bound violations, the nonrobust scheme fails in 171 of 192 time steps when more power than planned is generated and the storage is charged to maintain power balance.
Since SoC values larger than $1.0$ are physically invalid, the next MPC step in our study continues at $s_{\subsys{1}} = 1.0$.

It should be noted that power balance can be ensured in other ways than using the storage as a buffer. 
For instance, if power imports from and exports to the main grid are allowed at all times, using the grid as a buffer would not lead to bound violations as above. 
However, this can cause very high costs, for example, if electricity has to be imported in the evening at expensive prices. 
In contrast, the battery allows power to be stored until exports to the main grid become profitable. 
Indeed, over the entire period of two days, the adaptively robust NMPC scheme achieves total costs of $-5.2\cdot 10^{3}$ in microgrid \subsys{1} and thus makes profit despite the attack.
On the contrary, nonrobust NMPC causes total local costs of $2.3\cdot 10^{4}$, which is orders of magnitudes larger.
Considering that we aim for a strategy to increase the resilience of the system, which takes into account not only robustness but also performance in terms of induced costs, the battery as a buffer is therefore a reasonable choice that enables and favors high resilience.

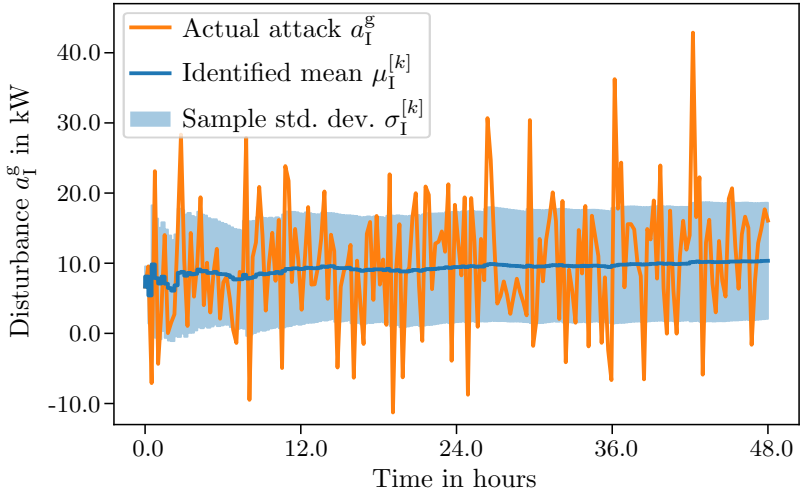
\begin{figure}[t!]%
	\hspace{1.7cm}%
	\begin{overpic}[width=0.8\linewidth]{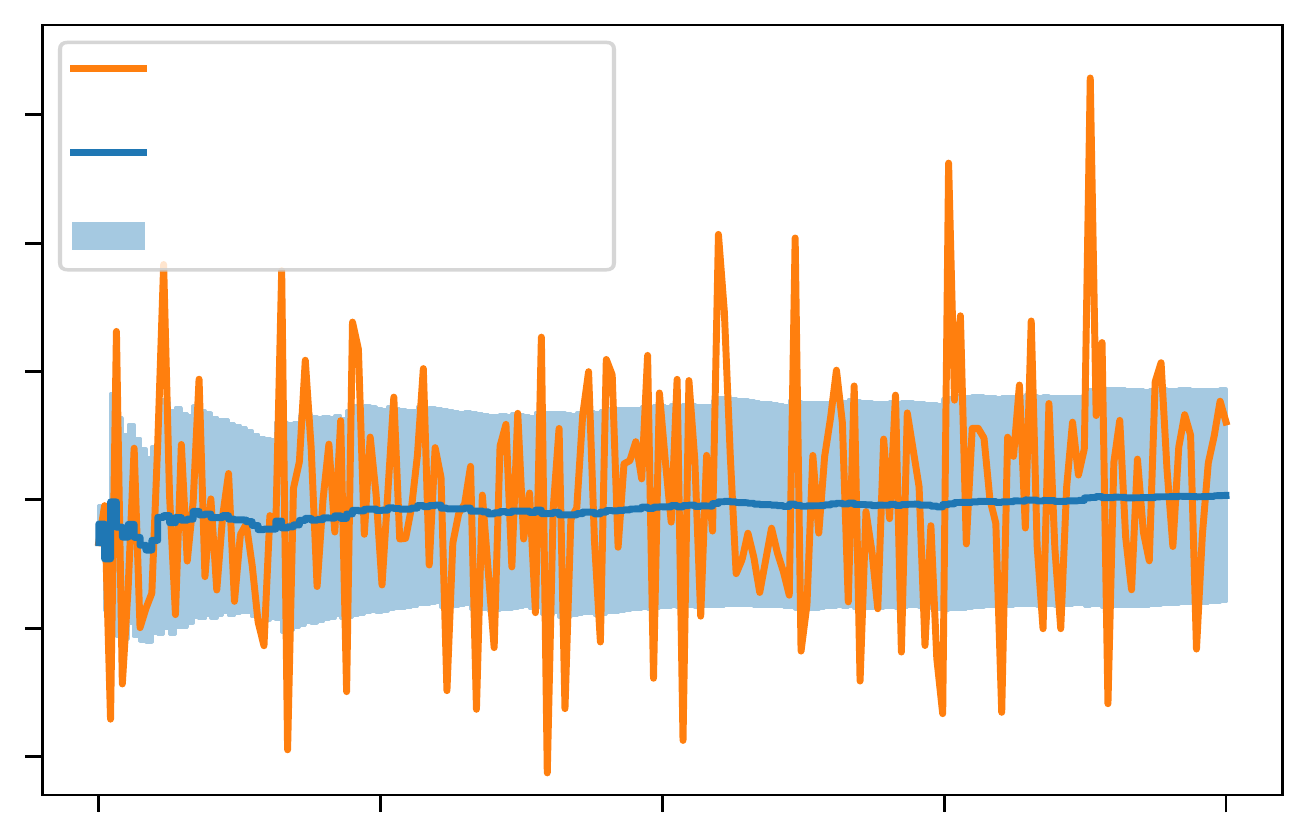}%
		% Legend
		\put(12.7, 57.5){Actual attack $a^{\text{g}}_{\subsys{1}}$}%
		\put(12.7, 51.25){Identified mean $\mu^{[k]}_{\subsys{1}}$}%
		\put(12.7, 44.9){Sample std.\ dev.\ $\sigma^{[k]}_{\subsys{1}}$}%
		% x-axis
		\put(39, -5.5){Time in hours}%
		\put(5.8, -1.2){\small 0.0}%
		\put(26.8, -1.2){\small 12.0}%
		\put(48.3, -1.2){\small 24.0}%
		\put(69.8, -1.2){\small 36.0}%
		\put(91.3, -1.2){\small 48.0}%		
		% y-axis
		\put(-11.5, 15.4){\rotatebox{90}{Disturbance $a^{\text{g}}_{\subsys{1}}$ in \si{\kilo\watt}}}%
		\put(-5.8, 54.3){\small \hphantom{-}40.0}%
		\put(-5.8, 44.4){\small \hphantom{-}30.0}%		
		\put(-5.8, 34.5){\small \hphantom{-}20.0}%
		\put(-5.8, 24.6){\small \hphantom{-}10.0}%
		\put(-5.8, 14.9){\small \hphantom{-1}0.0}%		
		\put(-5.8, 5.0){\small -10.0}%
	\end{overpic}%
	\vspace{0.8cm}%
	\caption{Course of the mean $\mu^{[k]}_{\subsys{1}}$ of identified values $a^{\ast,k}_{\subsys{1}}$ over time, with sample standard deviation $\sigma_{\subsys{1}}^{[k]}$. The actual disturbance $a_{\subsys{1}}^{\text{g},k}$ at each time k is shown in orange. The figure is taken from \cite[Fig.\ 3]{Braun2022Resilient}.}%
	\label{fig:microgrid adi results generation and renewables}%
\end{figure}

In the second experiment, we consider a modified generator attack \mbox{$a^{\text{g}}_{\subsys{1}} = 10.0\sinew{\kilo\watt} + r^{\text{g}}_{\subsys{1}}$}, where $r^{\text{g}}_{\subsys{1}} \sim \mathcal{N}(0.0, 8.0)\sinew{\kilo\watt}$ represents the uncertainty in renewable generation and is randomly drawn from a normal distribution with mean $0.0\sinew{\kilo\watt}$ and standard deviation~$8.0\sinew{\kilo\watt}$, independently at each time step. 
Together, the malicious attack of $10.0\sinew{\kilo\watt}$ and the renewable fluctuations $r^{\text{g}}_{\subsys{1}}$ may cause more power than planned to be generated (i.\,e., $a^{\text{g}}_{\subsys{1}} > 0$) or less (i.\,e., $a^{\text{g}}_{\subsys{1}} < 0$), but are chosen such that the total generator input $u^{\text{g}}_{\subsys{1}} + a^{\text{g}}_{\subsys{1}}$ is nonnegative. 
Due to the fluctuating generation, the actual value $a^{\text{g}}_{\subsys{1}}$ of the unknown disturbance in the generator dynamics ranges from $-11.3\sinew{\kilo\watt}$ to $42.9\sinew{\kilo\watt}$ as can be seen in \Cref{fig:microgrid adi results generation and renewables}.
For the examined generator with parameters as in \Cref{mpc:tab:microgrid parameter costs}, this is a very broad range, which also becomes clear in comparison with \Cref{fig:microgrid attack generation robust vs nonrobust trajectories:subfig:generation}.
As an apparent consequence of the continually changing values, the local identification problem \labelcref{opt:identification problem local} yields a different suspicion $a^{\text{g},\ast}_{\subsys{1}}$ in each time step.
Nevertheless, \Cref{fig:microgrid adi results generation and renewables} shows that the mean $\mu^{[k]}_{\subsys{1}}$ of identified values quickly settles at about $10.0\sinew{\kilo\watt}$, which underlines that the distributed ADI method is able to cope also with highly fluctuating and widely dispersed disturbances, since a new optimization problem is solved at each time step.
This proves once again the great potential of the proposed class of optimization-based ADI methods and emphasizes that they are not tailored to a specific type of attack, but are also very well suited for challenging scenarios where attacks and other sources of significant uncertainty congregate.

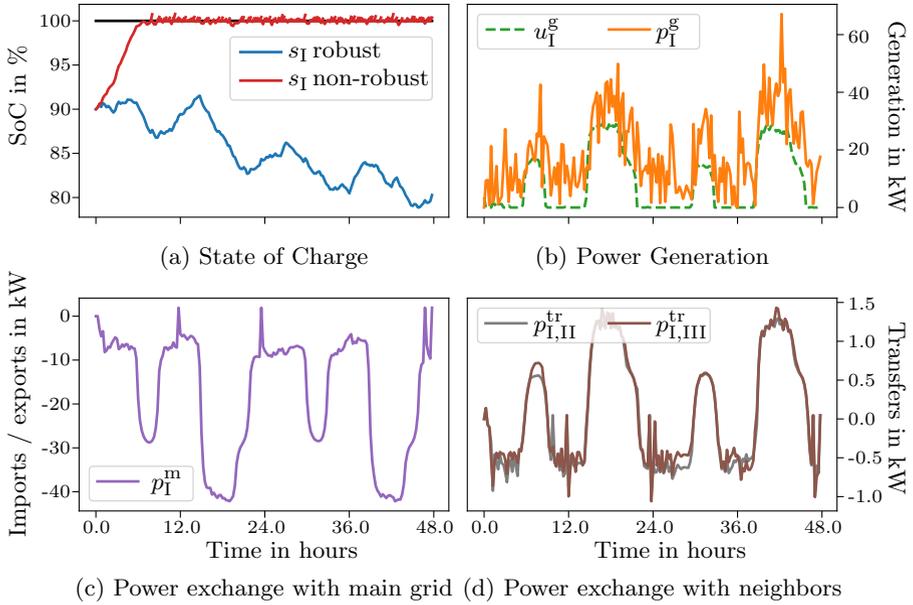
\begin{figure}[t!]%
	\hspace{0.9cm}%
	\begin{minipage}[c]{\linewidth}
		\begin{subfigure}{0.43\linewidth}%
			\begin{overpic}[width=\linewidth]{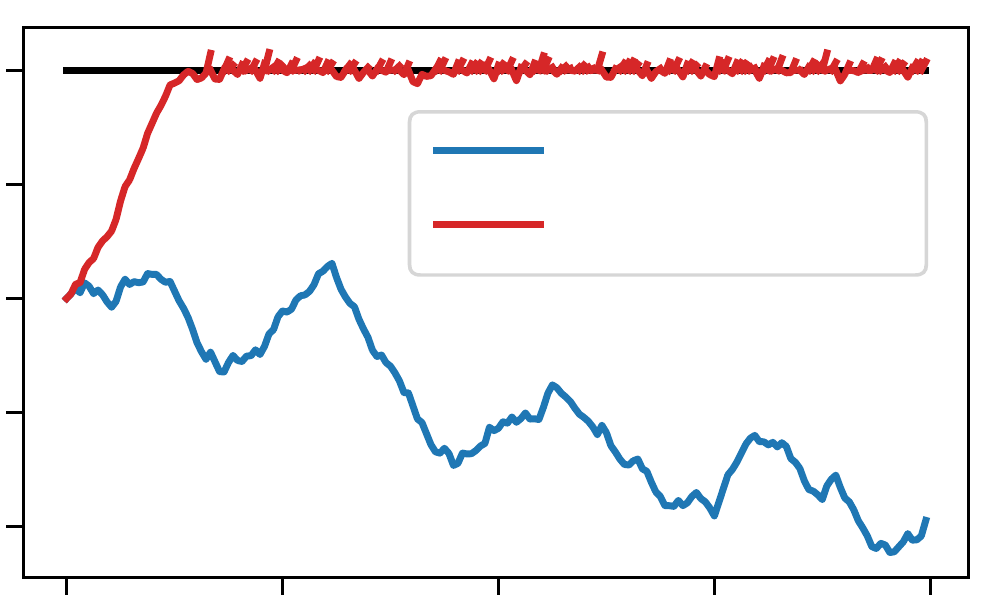}%
				% Legend entries
				\put(56.2, 44.4){\scriptsize $s_{\subsys{1}}$\,robust}
				\put(56.2, 37.0){\scriptsize $s_{\subsys{1}}$\,non-robust}
				% y-axis
				\put(-16, 21.4){\scriptsize \rotatebox{90}{SoC in \%}}%
				\put(-7.9, 52.4){\footnotesize 100}%		
				\put(-7.9, 40.9){\footnotesize \hphantom{1}95}%	
				\put(-7.9, 29.4){\footnotesize \hphantom{1}90}%				
				\put(-7.9, 17.9){\footnotesize \hphantom{1}85}%
				\put(-7.9, 6.4){\footnotesize \hphantom{1}80}%						
			\end{overpic}%
			\caption{State of Charge}%
		\end{subfigure}%
		\begin{subfigure}{0.43\linewidth}%
			\begin{overpic}[width=\linewidth]{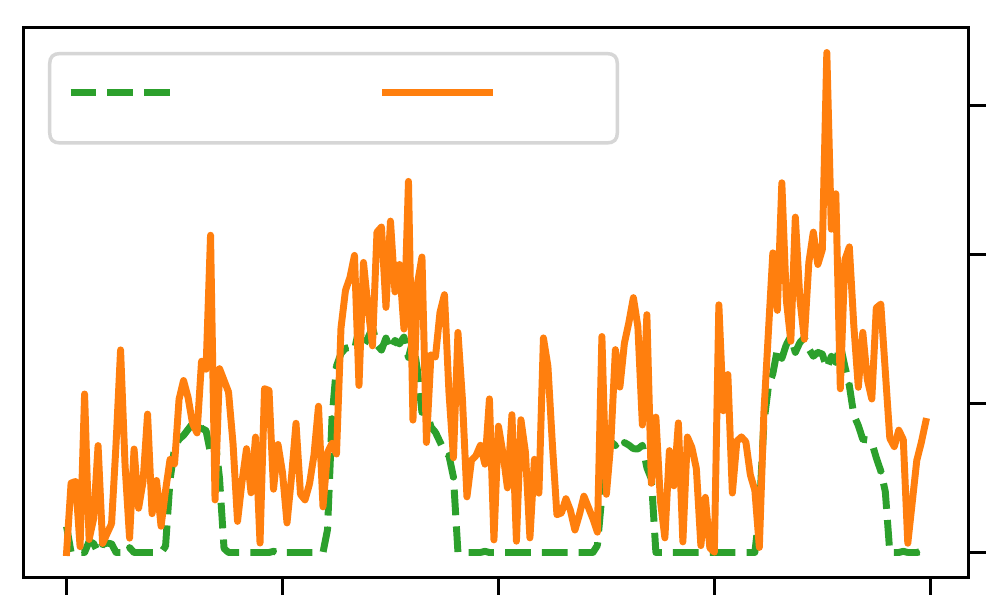}%
				% y-axis labels power generation	
				\put(111, 54){\scriptsize \rotatebox{-90}{Generation in \si{\kilo\watt}}}%	
				\put(100.5, 48.2){\footnotesize 60}%
				\put(100.5, 33.46){\footnotesize 40}%		    
				\put(100.5, 19.12){\footnotesize 20}%		    
				\put(100.5, 4.1){\footnotesize 0\hphantom{0}}%
				% Legend entries
				\put(19.5, 49.3){\scriptsize $u^{\text{g}}_{\subsys{1}}$}
				\put(51, 49.3){\scriptsize $p^{\text{g}}_{\subsys{1}}$}
			\end{overpic}%
			\caption{Power Generation}%
		\end{subfigure}%
		\newline%
		\begin{subfigure}{0.43\linewidth}%	
			\begin{overpic}[width=\linewidth]{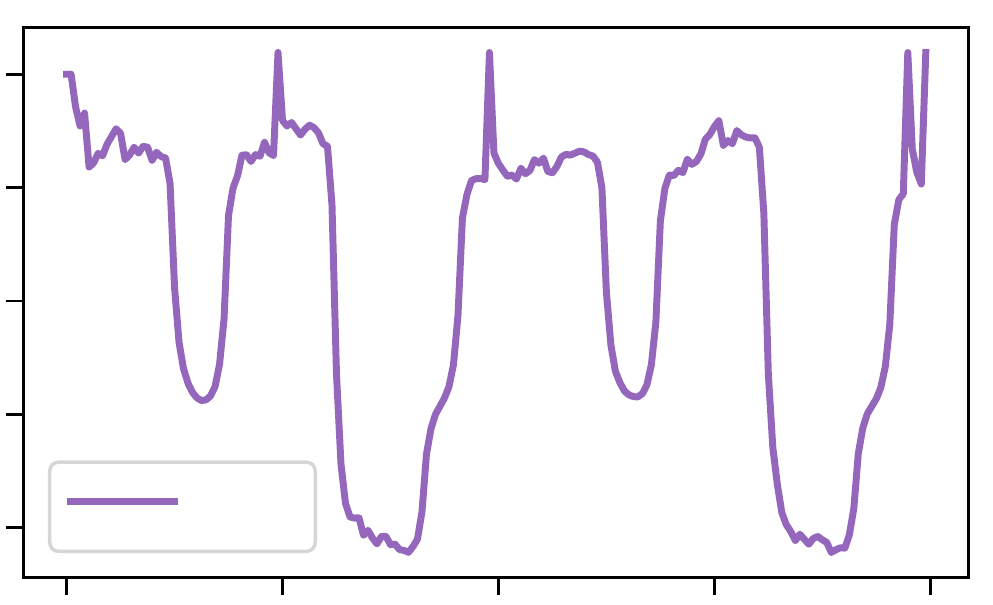}%
				% x-axis description and labels		
				\put(35, -9.4){\scriptsize Time in hours}%
				\put(3.4, -2.9){\footnotesize 0.0}%
				\put(24, -2.9){\footnotesize 12.0}%
				\put(45.8, -2.9){\footnotesize 24.0}%
				\put(67.5, -2.9){\footnotesize 36.0}%
				\put(89.2, -2.9){\footnotesize 48.0}%
				% y-axis labels
				\put(-16, -4){\scriptsize \rotatebox{90}{Imports / exports in \si{\kilo\watt}}}%			
				\put(-7.2, 52){\footnotesize \hphantom{-1}0}%
				\put(-7.2, 40.6){\footnotesize -10}%
				\put(-7.2, 29.2){\footnotesize -20}%		
				\put(-7.2, 17.7){\footnotesize -30}%
				\put(-7.2, 6.4){\footnotesize -40}%			
				% Legend entries
				\put(21., 8.5){\scriptsize $p^{\text{m}}_{\subsys{1}}$}%
			\end{overpic}%
			\vspace{0.5cm}%
			\caption{Power exchange with main grid}
		\end{subfigure}%
		\begin{subfigure}{0.43\linewidth}%
			\begin{overpic}[width=\linewidth]{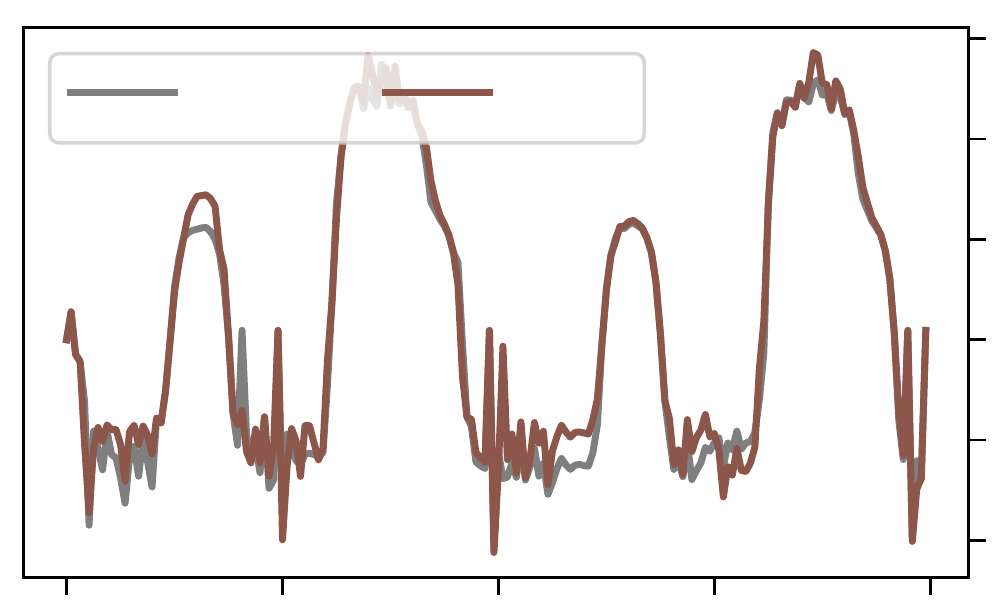}%
				% x-axis description and labels		
				\put(35, -9.4){\scriptsize Time in hours}%
				\put(3.4, -2.9){\footnotesize 0.0}%
				\put(24, -2.9){\footnotesize 12.0}%
				\put(45.8, -2.9){\footnotesize 24.0}%
				\put(67.5, -2.9){\footnotesize 36.0}%
				\put(89.2, -2.9){\footnotesize 48.0}%
				% y-axis
				\put(111, 52){\scriptsize\rotatebox{-90}{Transfers in \si{\kilo\watt}}}%
				% y-axis labels power transfer
				\put(100.5, 55.3){\footnotesize 1.5}%				
				\put(100.5, 45.25){\footnotesize 1.0}%				
				\put(100.5, 35.2){\footnotesize 0.5}%
				\put(100.5, 25.15){\footnotesize 0.0}%
				\put(100.5, 15.1){\footnotesize -0.5}%
				\put(100.5, 5.1){\footnotesize -1.0}%
				% Legend entries	
				\put(19.5, 49.5){\scriptsize $p^{\text{tr}}_{\subsys{1},\subsys{2}}$}%
				\put(51, 49.5){\scriptsize $p^{\text{tr}}_{\subsys{1},\subsys{3}}$}%
			\end{overpic}%
			\vspace{0.5cm}%
			\caption{Power exchange with neighbors}
		\end{subfigure}%
	\end{minipage}
	\caption{States and inputs in microgrid \subsys{1}, which now contains renewable generation as another source of uncertainty in addition to the generator attack.
		%  The figure is similar to \cite[Fig.\ 4]{Braun2022Resilient}.
	}
	\label{fig:microgrid attack generation and renewables robust vs nonrobust}%
\end{figure}

The sample standard deviation $\sigma^{[k]}_{\subsys{1}}$ is considerably larger than before and the three scenarios $\mu^{[k]}_{\subsys{1}}, \mu^{[k]}_{\subsys{1}} + \sigma^{[k]}_{\subsys{1}}$, and $\mu^{[k]}_{\subsys{1}} - \sigma^{[k]}_{\subsys{1}}$ are further apart than in the first experiment. 
\Cref{fig:microgrid attack generation and renewables robust vs nonrobust} shows the obtained solution for the attacked microgrid \subsys{1}.
While adaptively robust DMPC achieves total local costs of $3.1\cdot 10^3$ in microgrid~\subsys{1}, the nonrobust approach causes more than ten times higher total costs of $3.2\cdot 10^4$. 
Once again, classical nonrobust MPC proves to be unsuitable to control the disturbed system as it computes a solution that violates the upper bound of the state of charge in 113 of 192 time steps.

At this point, we would like to point out that the adaptively robust DMPC scheme is not guaranteed to yield admissible trajectories in all cases.
In fact, proving rigorous guarantees of this kind is challenging for nonlinear dynamics. 
Moreover, in contrast to the multi-stage approach \cite{Lucia2020Stability}, adaptively robust NMPC lacks the recursive feasibility property when the attack uncertainty sets $\mathcal{A}_I^{l, [k]}$ are adjusted to sudden attacks.
Furthermore, \Cref{fig:microgrid adi results generation and renewables} illustrates that in our second attack scenario involving uncertain renewable generation, even disturbances $a^{\text{g}}_{\subsys{1}}$ occur that are not within the interval $[\mu^{[k]}_{\subsys{1}} - \sigma^{[k]}_{\subsys{1}}, \mu^{[k]}_{\subsys{1}} + \sigma^{[k]}_{\subsys{1}}]$.
Despite these unforeseen disruptions and the lack of theoretical guarantees, however, all state bounds are satisfied and the solution in \Cref{fig:microgrid attack generation and renewables robust vs nonrobust} is not overly conservative judging from the fact that considerably lower costs are obtained than with nonrobust DMPC.
This underlines that adaptively robust NMPC, using ADI results as estimates for an unknown attack, is a very powerful tool even under challenging circumstances with broadly dispersed disturbances.

\section{Conclusion and Future Directions}
\label{sec:conclusion}
We introduced a comprehensive distributed MPC framework for nonlinear control systems under attack, which is based on local multi-stage control and novel distributed attack identification methods in each subsystem.
To enable the system to respond autonomously and robustly to identified perturbations, each control scheme represents the uncertain influence of neighboring couplings and attack inputs by scenario sets that are continuously updated based on newly gained knowledge.
For this purpose, each subsystem applies local attack identification and repeatedly transmits new contract information to its neighbors.
Using the example of microgrids interconnected by power transfers, the methodology was demonstrated to robustly control a distributed system and achieve constraint satisfaction at all times despite unknown attacks and uncertain renewable generation.

We have identified two promising directions with great potential for future research.
The first would be to derive theoretical conditions under which \Cref{algo:distributed adi} can be rigorously proven to successfully identify the correct inputs, similar to the guarantees for our centralized ADI method \cite{Braun2021Attack}.
While some ideas from \cite{Braun2021Attack} can be transferred with few changes, further required theoretical arguments could be based on the research results on \emph{nonlinear} compressed sensing. 
For example, in \cite{Blumensath2013Compressed} the restricted isometry property from \cite{Candes2005Decoding}, a central component of linear compressed sensing, is generalized and the iterative hard thresholding algorithm  involving a form of gradient projection is extended to nonlinear systems.
Furthermore, in \cite{Beck2013Sparsity} two coordinate descent methods are introduced that build upon the simplex algorithm for linear programming and are of a greedy type in the sense that they add nonzero variables one by one.
When suitable success guarantees for the new distributed ADI approaches provably hold, a combination with the robustness and stability analysis of multi-stage NMPC and contract-based DMPC described in \cite{Lucia2015Contract,Lucia2020Stability,Lucia2014Multi} could be the next step to strengthen the excellent numerical performance of adaptively robust DMPC by theoretical arguments.

The second research direction consists in investigating a hierarchical combination of several ADI approaches that complement each other and provide system operators with different options suiting their needs.
There is, on the one hand, the centralized ADI method from \cite{Braun2021Attack}, which is based on an approximation of the dynamics and provides quick insights into the network-wide attack situation, but requires all subsystems to make specific sensitivity information publicly available and agree on a central instance to solve the global identification problem.
On the other hand, there are distributed ADI methods like \Cref{algo:distributed adi} involving problems \labelcref{opt:identification problem local,opt:identification problem distributed}, which use local models to analyze possible attacks on one subsystems or its neighborhood locally.
Several gradations or variants of these approaches may be applied, depending on the available model knowledge and the willingness of individual subsystems to cooperate or agree on a common decision instance.

\section{Statement on Conflict of Interests}
On behalf of all authors, the corresponding author states that there is no conflict of interest.

\bibliography{literature_springer_natural}% common bib file
\bibliographystyle{sn-mathphys}%
%%\input main.bbl

\end{document}